\documentclass[12pt]{article}
\pdfoutput=1
\usepackage[top=30truemm,bottom=30truemm,left=25truemm,right=25truemm]{geometry}
\usepackage{authblk}
\usepackage{amsthm}
\usepackage{amsmath,amssymb,bm,mathtools}
\usepackage{cite}
\usepackage[]{hyperref}
\hypersetup{colorlinks,linkcolor={blue},citecolor={blue},urlcolor={blue}}  
\usepackage[titletoc]{appendix}
\usepackage{url}
\usepackage{graphicx,color}
\usepackage{braket}
\usepackage{qcircuit}
\usepackage{here}
\usepackage{mathrsfs}
\usepackage{subcaption}
\usepackage{tikz}
\usepackage{titlesec}

\titleformat*{\section}{\large\bfseries}

\newtheorem*{definition*}{Definition}
\newtheorem*{assumption*}{Assumption}

\newcommand{\red}[1]{\color{red}{#1}}

\def\sinh{{\mathrm{sinh}}}
\def\cosh{{\mathrm{cosh}}}
\def\tanh{{\mathrm{tanh}}}

\def\t{{\theta}}

\def\be{\begin{equation}}
\def\ee{\end{equation}}
\def\ba{\begin{eqnarray}}
\def\ea{\end{eqnarray}}

\numberwithin{equation}{section}

\def\Z2{\mathbb{Z}_2}

\AtBeginDocument{%
  {\footnotesize\global\footnotesep=0.8\baselineskip}%
}

\definecolor{cardinal}{rgb}{0.6,0,0}
\definecolor{darkgreen}{rgb}{0,0.4,0}
\definecolor{golden}{rgb}{0.92, 0.7, 0}
\definecolor{midnight}{rgb}{0, 0, 0.5}
\definecolor{darkblue}{rgb}{0, 0, 0.7}
\definecolor{purple}{rgb}{0.5, 0, 0.5}

\usepackage{tikz}
\usetikzlibrary{decorations.pathreplacing}

\begin{document}

\begin{titlepage}
\thispagestyle{empty}

\begin{flushright}
\end{flushright}

\bigskip

\begin{center}
\noindent{\bf \Large No boundary density matrix in elliptic de Sitter dS/$\mathbb{Z}_2$}\\


\vspace{1.4cm}

{\bf Rapha\"el Dulac$^	\natural
$ and Zixia Wei$^\sharp$}
\vspace{1cm}\\

{\it
$^\natural$Institut de Physique Th\'eorique,
Universit\'e Paris Saclay, 
CEA, CNRS, 
Orme des Merisiers, Gif sur Yvette, 91191 CEDEX, France
}\\[1.5mm]

{\it 
$^\sharp$Society of Fellows, 
Harvard University, Cambridge, MA 02138, USA
}\\[1.5mm]



\vspace{0.3cm}


\medskip

\end{center}

\begin{abstract}

Elliptic de Sitter (dS) spacetime dS$/\mathbb{Z}_2$ is a non-time-orientable spacetime obtained by imposing an antipodal identification to global dS. 
Unlike QFT on global dS, whose vacuum state can be prepared by a no-boundary Euclidean path integral, the Euclidean elliptic dS does not define a wavefunction in the usual sense. We propose instead that the path integral on the Euclidean elliptic dS defines a no-boundary density matrix. As an explicit example, we study the free Dirac fermion CFT in two-dimensional elliptic dS and analytically compute the von Neumann and the R\'enyi entropies of this density matrix. The calculation reduces to correlation functions of vertex operators on non-orientable surfaces. As a by-product, we compute the time evolution of entanglement entropy following a crosscap quench in free Dirac fermion CFT. We also comment on a striking feature of free QFT in elliptic dS: its global Hilbert space is one-dimensional, wheres the Hilbert space associated to each observer is a nontrivial Fock space.

\end{abstract}

\end{titlepage}

\newpage
\setcounter{page}{1}
\tableofcontents

\section{Introduction}

The de Sitter (dS) spacetime is the maximally symmetric solution of the Einstein equations with a positive cosmological constant. It is of great interest from the phenomenological point of view since our universe is presently dominated by dark energy and approximately dS \cite{Planck18}. Besides, the very early stage of our universe is also very close to the dS geometry due to the exponential expansion driven by the inflaton potential \cite{Guth80,Linde81}. Therefore, understanding the quantum effects in dS is crucial for solving the cosmological problems in our own universe. 

On the other hand, developing a quantum theory of gravity in dS is also of great theoretical interests. One of the most successful approaches of quantum gravity is the holographic principle \cite{tHooft93,Susskind94}, and the AdS/CFT correspondence \cite{Maldacena97} serves as the most well-understood example. The AdS/CFT correspondence states that a quantum gravitational theory in $(d+1)$-dimensional asymptotically anti-de Sitter (AdS) spacetime is equivalent to a $d$-dimensional conformal field theory (CFT), allowing us to understand mysterious quantum gravity through the lens of a well-defined non-gravitational quantum theory non-perturbatively.  
However, a complete theory of quantum gravity should be applicable in any spacetime. dS serves as a canonical playground for finding such a theory. Among such attempts, variant ideas of how the holographic principle should be formulated for dS gravity, such as the dS/CFT correspondence \cite{Strominger01,Hull:1998vg,Maldacena02,AHS11,HNTT21}, the DS/dS correspondence \cite{AKST04,DST18}, and different types of static patch holography \cite{Susskind21-0, Susskind21,Susskind21-2,NV23,Verlinde24,VZ24}, have been proposed. 

\begin{figure}[t]      \centering\includegraphics[width=13cm]{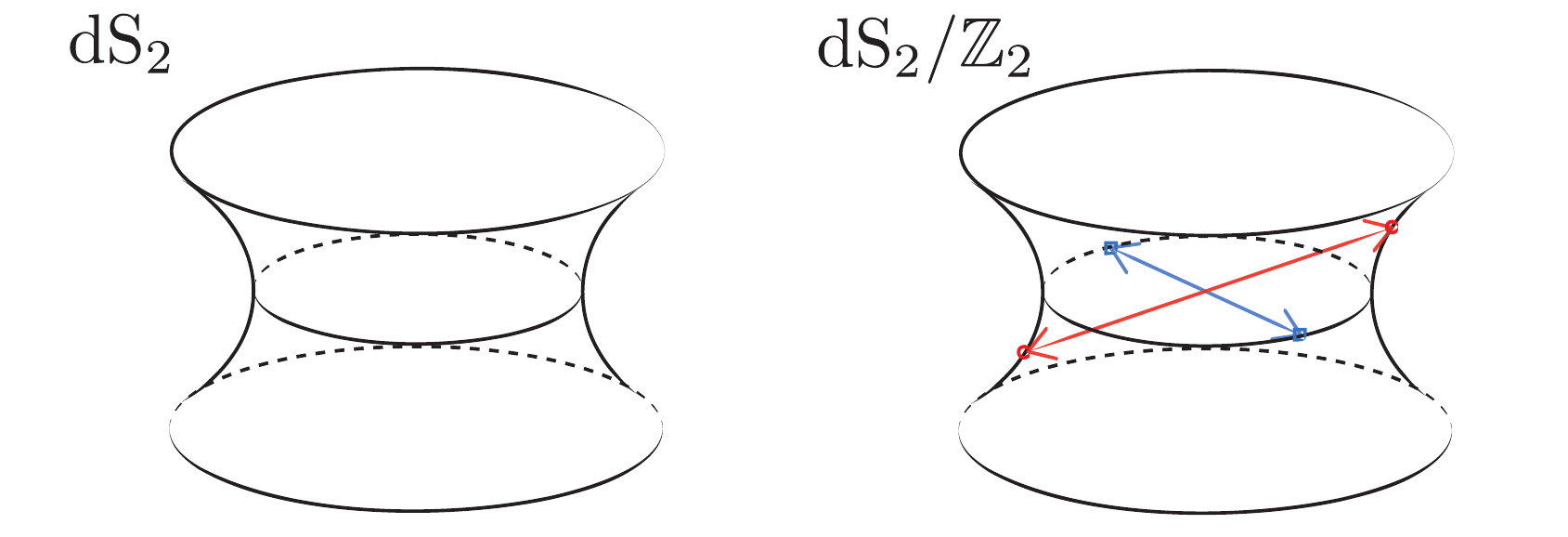}
    \caption{A sketch of global dS$_2$ and elliptic dS$_2$.}
    \label{fig:dS_EdS}
\end{figure}

Meanwhile, there exists another solution to the Einstein equations with a positive cosmological constant which is locally maximally symmetric, known as the elliptic dS. This is a solution obtained by a $\mathbb{Z}_2$ quotient of global dS identifying the antipodal points. 
See figure \ref{fig:dS_EdS} for a sketch. 
This $\Z2$ identification firsly appears in Schr\"odinger's book\cite{Schrodinger56} in 1956 where he speculated that such dS$/\Z2$ might be a better spacetime compared to global dS. The basic properties of dS$/\Z2$ were later invesigated by \cite{PSV02}. 

In the traditional approach of constructing dS holography, one usually starts from the dS spacetime, place a holographic screen in it, and then try to figure out the dual theory living on the holographic screen. 
An alternative but more promising approach is to start from a non-gravitational quantum theory under better control, and then try to find dS features emerging from it. 
In some recent proposals of bottom-up dS holography under such philosophy \cite{Wei24,BTV25}, one finds it is the Euclidean counterpart of elliptic dS which naturally emerges. It is surprising such a feature appears in different proposals, and this motivates us to understand the physical meaning of the Euclidean quantum field theory (QFT) on Euclidean elliptic dS is.

\begin{figure}[t]      \centering\includegraphics[width=13cm]{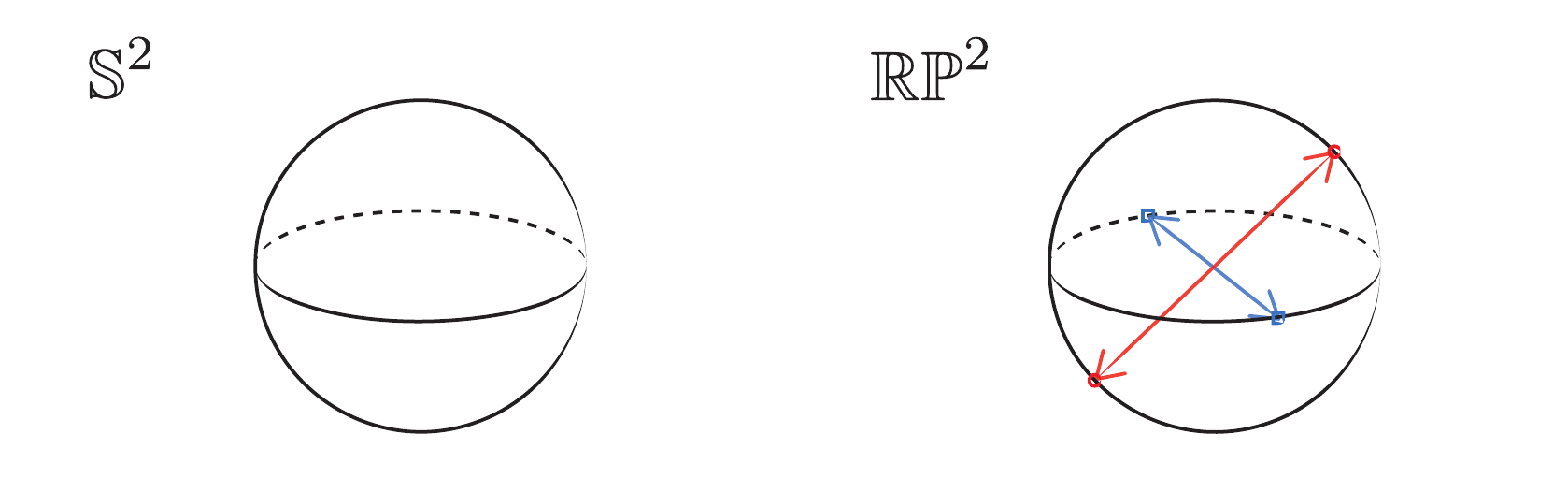}
    \caption{A sketch of $\mathbb{S}^2$ and $\mathbb{RP}^2$. They are the Euclidean counterparts of dS$_2$ and dS$_2/\mathbb{Z}_2$. }
    \label{fig:S_RP2}
\end{figure}

The situation is, however, very different from the relation between global dS$_{d+1}$ and its Euclidean counterpart $\mathbb{S}^{d+1}$. A Euclidean path integral over $\mathbb{S}^{d+1}$ can be split into two hemispheres in a time-reflection symmetric way, where one computes the no-boundary wave function $\ket{\Psi_{\rm NB}}$ \cite{BD78,HH83} defined on the equator $\mathbb{S}^{d}$ and the other computes $\bra{\Psi_{\rm NB}}$. Therefore, real time correlators on dS$_{d+1}$ obtained by performing analytic continuation of Euclidean correlators on $\mathbb{S}^{d+1}$ computes the correlators on the time evolved no-boundary wave function. On the other hand, the Euclidean counterpart of dS$_{d+1}/\Z2$ is the real projective space $\mathbb{RP}^{d+1} = \mathbb{S}^{d+1}/\mathbb{Z}_2$, which cannot be split into two parts in a time-reflection symmetric way. See figure \ref{fig:S_RP2} for a sketch. Therefore, path integral over $\mathbb{RP}^{d+1}$ cannot be interpreted as a wave function, which makes the state interpretation of the correlators computed from the analytic continuation of $\mathbb{RP}^{d+1}$ correlators unclear. 

Despite the fact that the Euclidean QFT on $\mathbb{RP}^{d+1}$ cannot be interpreted as a wave function on a spatial slice, in this paper, we propose that one can associate an interpretation of no-boundary density matrix to it. 
The original idea of the no-boundary density matrix is a generalization of Hartle-Hawking's (HH) no-boundary wave function recently proposed by Ivo, Li and Maldacena (ILM) in \cite{ILM24} to compute a density matrix on a subregion in gravitational path integral. As opposed to the HH proposal where one fixes the data on a Cauchy slice and sums over all the possible geometries and field configurations consistent with it, the ILM proposal only imposes boundary conditions on a spatial subregion. In this paper, we are considering a purely field theoretical analogue of the ILM no-boundary density matrix on a fixed background geometry dS$/\mathbb{Z}_2$ where the no-boundary wave function cannot be defined. No dynamical gravity is involved here. The details of this proposal will be described in section \ref{sec:EdS_NBDM}, after a summary of basic properties of the dS$/\mathbb{Z}_2$. 

As an explicit example, we study the free Dirac fermion conformal field theory (CFT) on dS$/\Z2$. The spectrum of such a density matrix can be fully characterized using its entanglement entropy and R\'enyi entropy. This is technically reduced to the problem of computing the replica partition function \cite{CC04} in crosscap CFTs (XCFT). The methodology is developed in section \ref{sec:replica_trick_CFT}, and the entropies will be computed in section \ref{sec:EERP2}.

Although the main topic in this paper is the Euclidean path integral on Euclidean elliptic dS, in section \ref{sec:1D_space}, we revisit the direct canonical quantization in the Lorentzian signature. We will emphasize its relationship with our Euclidean analysis, and point out an interesting feature of QFT on dS$/\mathbb{Z}_2$ ---  its global Hilbert space is one-dimensional, but a nontrivial Hilbert space associated to each observer can be defined. We will close with some conclusions and remarks in section \ref{sec:conclusion}.

In appendix \ref{app:A}, we list out the explicit expressions for the 2-point function of twist operators on non-orientable surfaces in different cases. In appendix \ref{app:quench}, we apply some results developed in the main text to study another physical setup, the crosscap quench \cite{CCR24,WY24}.

\section{Elliptic de Sitter and no-boundary density matrix}\label{sec:EdS_NBDM}
In this section, we first define the elliptic dS and study its causal structure. Then we describe the proposal that the Euclidean QFT on dS$/\mathbb{Z}_2$ can be interpreted as a no-boundary density matrix, and explain how to compute the entropies that characterize its spectrum.

\subsection{Definition and causal structure}\label{sec:causal_structure}
\begin{figure}[h]      \centering\includegraphics[width=16cm]{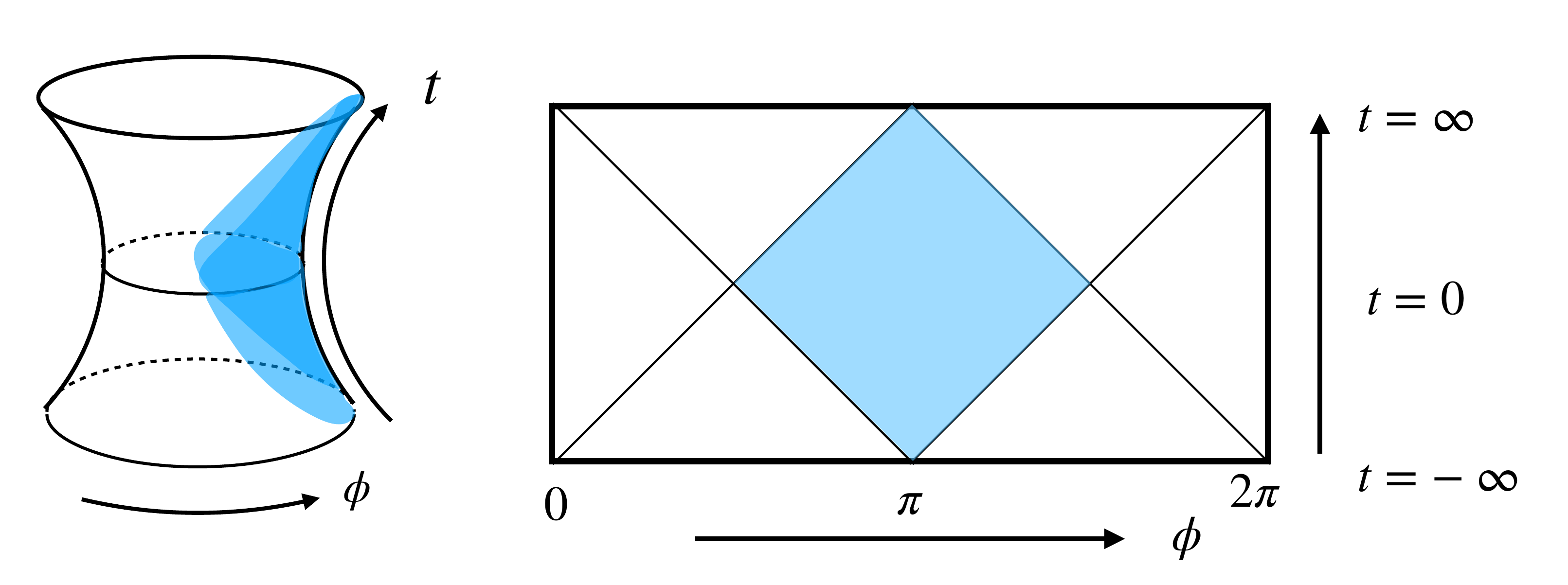}
    \caption{A sketch of global dS$_2$ with metric $ds^2 = L_{\rm dS}^2 \left( -dt^2 + \cosh(t)^2d\phi^2\right )$ where $t\in (-\infty,\infty)$ and $\phi \in (-\pi,\pi]$ with $\phi \sim \phi + 2\pi$ is shown. The right panel shows its Penrose diagram. The static patch associated to the static observer sitting at $\phi =0$ is shaded. The static patch can be expressed as $|\phi| + \arcsin(\tanh |t|) < \pi/2$. Its boundary is called the cosmological horizon.}
    \label{fig:dS2}
\end{figure}

Consider the $(d+2)$-D Minkowski spacetime with the metric 
\begin{align}
    ds^2 = -dX_0^2 + dX_1^2 + \cdots +dX_{d+1}^2. 
\end{align}
The $(d+1)$-D global dS spacetime dS$_{d+1}$ is defined as the codimension-1 hypersurface on which the distance from each point to the origin is equal: 
\begin{align}
    -X_0^2 + X_1^2 + \cdots +X_{d+1}^2 = L_{\rm dS}^2,
\end{align}
and $L_{\rm dS}$ is called the de Sitter radius. From this definition, it manifests that dS$_{d+1}$ has an $O(d+1,1)$ isometry group. For $d\geq 2$, dS$_{d+1}$ is the maximally symmetric solution to the Einstein equations
\begin{align}
    G_{\mu\nu} + \Lambda g_{\mu\nu} = 0,  
\end{align}
with a positive cosmological constant $\Lambda$, which is related to $L_{\rm dS}$ as
\begin{align}
    \Lambda = \frac{d(d-1)}{2L_{\rm dS}^2}.
\end{align}
The metric of dS$_{d+1}$ can be written intrinsically as 
\begin{align}\label{eq:dSmetric}
    ds^2 = L_{\rm dS}^2 \left( -dt^2 + \cosh^2 t ~d\Omega_{d}^2\right ), 
\end{align}
where $d\Omega_{d}^2$ is the line element of a $d$-D unit sphere $\mathbb{S}^{d}$. This is the so-called the global coordinate of dS since it covers the maximal extension of it. Each $t={\rm const.}$ slice is a $\mathbb{S}^{d}$. The topology of global dS is $\mathbb{S}^{d} \times \mathbb{R}$. 
Consider a static observer $O$ sitting at fixed $\Omega_d$ and traveling between $t\in (-\infty,\infty)$. 
The causal diamond associated with such an observer, which does not cover the whole dS$_{d+1}$, is called a static patch of global dS. In figure \ref{fig:dS2}, a sketch of global dS$_2$ and one of its static patch are shown. 
The null boundaries of the static patch are called the cosmological horizon, outside of which the observer $O$ cannot communicate with. A review of dS can be found in \cite{SSV01}. 

The elliptic de Sitter is defined as a $\Z2$ quotient of global dS$_{d+1}$ by imposing the identification \cite{Schrodinger56, PSV02}
\begin{align}
    (X_0, X_1,\cdots, X_{d+1}) \sim (-X_0, -X_1,\cdots, -X_{d+1}). 
\end{align}
Note that this is a spacetime antipodal identification, which manifests in the global coordinate as $(t, \Omega_d) \sim (-t, \Omega_d^A)$, where $\Omega_d^A$ denotes the spatial antipodal point of $\Omega_d$ on the unit sphere $\mathbb{S}^d$. 
The topology of the $t=0$ slice is $\mathbb{RP}^d$, and the other $t={\rm const.}\neq 0$ slices are $\mathbb{S}^d$. Therefore, there is no continuous time evolution which brings the $t=0$ slice to a $t={\rm const.}\neq 0$ slice because of the topological obstruction. 

Since the antipodal identification inverts the time direction, one may worry if the causal structure of dS$/\Z2$ is pathological. Let us summarize several important features of the causality in dS$/\Z2$, some of which has been previously studied in \cite{Schrodinger56, PSV02}. 
Indeed, dS$/\Z2$ does not admit a global time orientation. However, due to the exponential expansion nature, there does not exist closed causal curves in it.\footnote{If one were to include the asymptotic infinity of dS$/\Z2$ and consider its conformal compactification, then there would be closed null curves passing through the asymptotic infinity.} This feature results in a nice causal structure for the physical events seen by a local observer. 

\begin{figure}[h]      \centering\includegraphics[width=15cm]{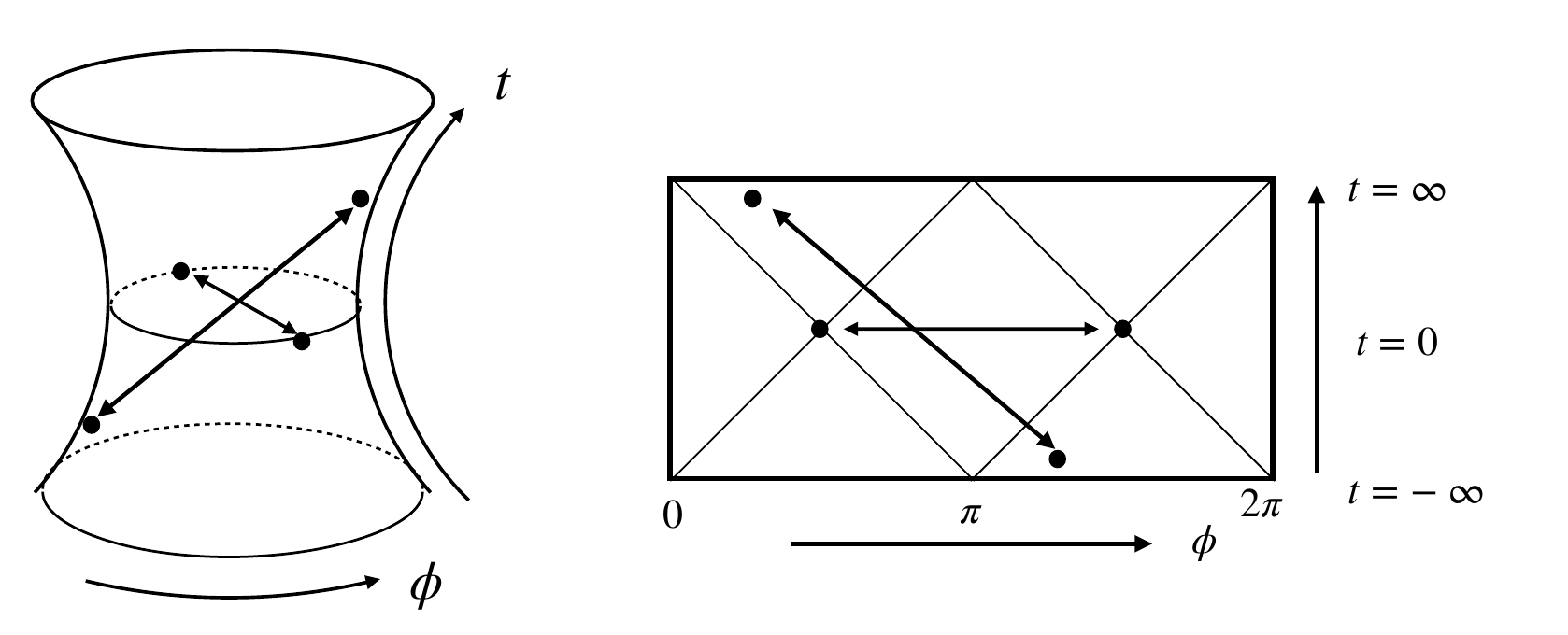}
    \caption{The left panel shows how a 2D elliptic de Sitter is obtained from a global dS$_2$ via the  $\mathbb{Z}_2$ quotient $(t,\phi) \sim (-t, \phi+\pi)$. The right panel shows how this is reflected in the Penrose diagram.}
    \label{fig:dS2_Z2}
\end{figure}

Consider a static observer $O$ traveling along a timelike trajectory specified by fixed $\Omega_d$. Such a timelike trajectory extends between two points at the asymptotic infinity which we may call $p$ and $q$. 
This timelike trajectory has the topology $\mathbb{R}$,
Therefore, we can use the ``global time" $t\in(-\infty,\infty)$ to parameterize it and associate it a time orientation. Let us identify $p$ ($q$) as its past (future) infinity which is approached at $t\rightarrow -\infty$ ($t\rightarrow +\infty$). figure~\ref{fig:dS2_Z2} shows a sketch of dS$_2/\mathbb{Z}_2$ and the world line an observer sitting at $\phi = 0$. The time orientation associated to the world line of $O$ can be consistently extended to the their ``static patch". In the following, the words ``past" and ``future" will be used with respect to this time orientation associated to the observer $O$.

\begin{figure}[h]      \centering\includegraphics[width=14cm]{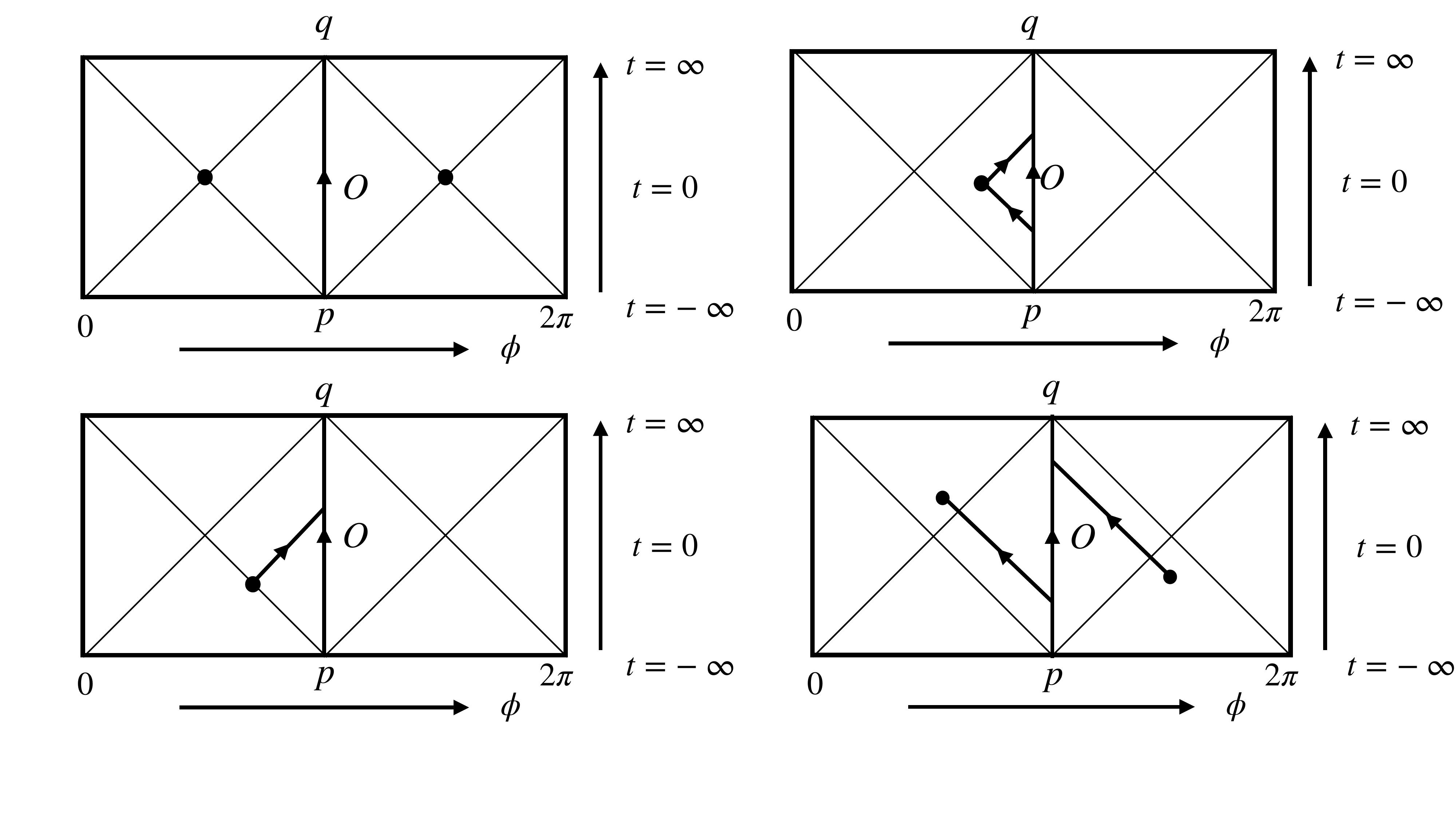}
    \caption{The upper left panel shows that a spacetime point living on the intersection of the light cones of $p$ and $q$ can neither send a signal to $O$ or receive a signal from them. The lower left panel shows that a spacetime point living on the light cone of $p$ ($q$) but not that of $q$ ($p$) can send a signal to (receive a signal from) $O$ but not receive a signal from (send a signal to) them. All other points can communicate with $O$. The case of points in the static patch is shown in the upper right panel, and the case of the other points is shown in the lower right panel. Note that the Penrose diagrams are shown in the unfolded way. }
    \label{fig:dS2_Z2_causality}
\end{figure}

It is straightforward to figure out that the observer $O$ can communicate with (i.e. can both send a signal to and receive a signal from\footnote{More precisely, we say that $O$ can send a signal to (receive a signal from) a spacetime point $r$ if there exists a point $s$ on the world line of $O$ such that there exists a null ray connecting $s$ and $r$ and $s\rightarrow r$ points towards the future (past) with respect to the time orientation inside the static patch.}) almost all the spacetime points in dS$/\Z2$. To see this, let us classify the spacetime points into several different classes. 
Refer to figure~\ref{fig:dS2_Z2_causality} for a sketch in dS$_2/\mathbb{Z}_2$.  
First of all, a point sitting on the intersection of the light cone of $p$ and that of $q$ can neither receive a signal from the observer $O$ nor send a signal to $O$. Secondly, a point sitting on the light cone of $p$ but not on the light cone of $q$ can send a signal to $O$ but not receive a signal from $O$. Similarly, a point sitting on the light cone of $q$ but not on the light cone of $p$ can receive a signal from $O$ but not send a signal to $O$. All the other points can communicate with $O$, i.e. both send a signal to $O$ and receive a signal from $O$, but these points can also be classified into two types. The first type is the ones inside the ``static patch", which can obviously communicate with $O$. The second type is the ones outside of the ``static patch". 
Consider such a point $r$, then one of its antipodal pair counterpart in global dS before the $\mathbb{Z}_2$ identification can send a signal to $O$ but not receive a signal from $O$, and the other one the opposite. Furthermore, it is worthy noticing that the timing when $O$ can receive a signal from a spacetime point is always in the future of the timing when $O$ can send a signal to that spacetime point, in both cases. Therefore, $O$ observes events in a consistent time ordering. 

In short, although the dS$/\mathbb{Z}_2$ itself does not admit a global time orientation, a static observer living in it observes a consistent time ordering and cannot figure out the spacetime is not time-orientable. Such an observer has causal access to almost all the spacetime, but not actually all. This fact will become important in the discussions later. 

\subsection{Euclidean elliptic de Sitter and no-boundary density matrix}
The Euclidean counterpart of global dS$_{d+1}$ is the $(d+1)$-D sphere $\mathbb{S}^{d+1}$, whose metric can be written by performing the analytic continuation $\theta = it$ in \eqref{eq:dSmetric}, 
\begin{align}\label{eq:RP2metric}
    ds^2 = L_{\rm dS}^2 \left(d\theta^2 +\cos^2\theta~ d\Omega_d^2\right). 
\end{align}
Here, $\theta \in \left[-\frac{\pi}{2},\frac{\pi}{2}\right]$. Accordingly, the Euclidean counterpart of elliptic de Sitter dS$_{d+1}/\mathbb{Z}_2$ is given by the antipodal identification $(\theta, \Omega_d) \sim (-\theta, \Omega_d^A)$ of $\mathbb{S}^{d+1}$, namely the $(d+1)$-D real projective space $\mathbb{RP}^{d+1}$. 

In this paper, we would like to consider what role is played by a Euclidean QFT on $\mathbb{RP}^{d+1}$ when understanding Lorentzian QFT on dS$_{d+1}/\mathbb{Z}_2$. In order to do that, let us first review what role is played by a Euclidean QFT on $\mathbb{S}^{d+1}$ when understanding Lorentzian QFT on global dS$_{d+1}$, and why its direct analogy is not applicable to the dS$/\mathbb{Z}_2$ case. After that, we will introduce our proposal. In the following, we consider the $d=1$ case just for simplicity. All the discussions can be applied directly to higher dimensional situations. 

\begin{figure}[htbp]      \centering\includegraphics[width=15cm]{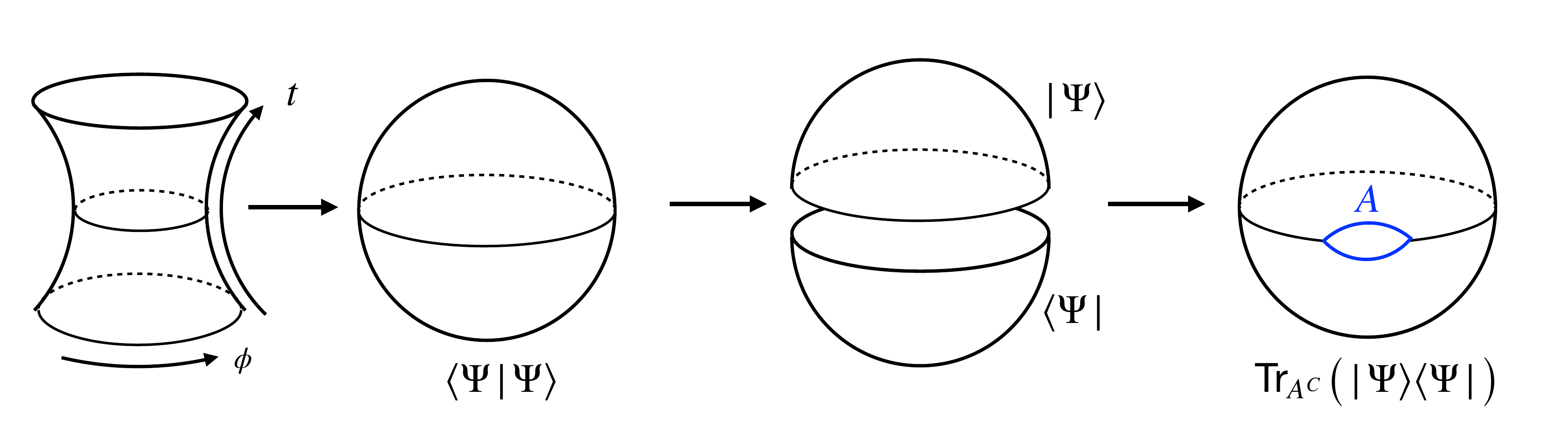}
    \caption{The Euclidean counterpart of dS$_2$ is $\mathbb{S}^2$, which admits $\theta \rightarrow -\theta$ as a time-reflection symmetry whose fixed locus separates it into two hemispheres. The path integral over a hemisphere prepares a no-boundary state $\ket{\Psi}$ on $\theta=0$. The sphere partition function computes $Z_{\mathbb{S}^2} = \braket{\Psi|\Psi}$. The path integral over the sphere with an open slit at $\theta=0$ along the subsystem $A$ computes ${\rm Tr}_{A^C}|\Psi\rangle\langle\Psi|$, where $A^C$ is the complement of $A$.}
    \label{fig:QFT_S2} 
\end{figure}

Let us firstly review the case in global dS, which is sketched in figure \ref{fig:QFT_S2}.  
When considering QFT on global dS$_{2}$, besides performing the canonical quantization and constructing Hilbert spaces and vacuum states directly in the Lorentzian signature \cite{CT68,Mottola84,Allen85,SSV01,BMS01}, we can also consider preparing a vacuum state using a Euclidean path integral over a hemisphere \cite{BD78,HH83,Miller25}. Consider a hemisphere given by $\theta \in \left[-\frac{\pi}{2},0 \right]$ and a field $\varphi(\theta,\phi)$ living on it. Consider a field theory whose Lagrangian density is given by $\mathcal{L}[\varphi(\theta,\phi)]$. Impose a boundary condition $\varphi(0,\phi) = \tilde{\varphi}(\phi)$ and perform the path integral over the hemisphere, we obtain a functional 
\begin{align}
    \Psi[\tilde{\varphi}(\phi)] \equiv \int_{\varphi(0,\phi) = \tilde{\varphi}(\phi)} \mathcal{D}\varphi ~\exp\left(-\int_{-\frac{\pi}{2}}^0 L_{\rm dS} ~\sin\theta~d\theta \int_{-\pi}^{\pi} d\phi ~\mathcal{L}[\varphi(\theta,\phi)]   \right). 
\end{align}
This is identified as the wave functional of a state 
\begin{align}
    \ket{\Psi} \equiv \int \mathcal{D}\tilde{\varphi} ~\ket{\tilde{\varphi}} \Psi[\tilde{\varphi}(\phi)], 
\end{align}
in the QFT Hilbert space defined on the $\theta=0$ circle with length $2\pi L_{\rm dS}$.
This state is often called the no-boundary state\footnote{It is also called the Euclidean vacuum state, the Bunch-Davis state \cite{BD78} or the Hartle-Hawking state \cite{HH83}.}, in the sense that it is prepared by a path integral in which no additional boundary conditions needs to be imposed other than on the slice $\theta = 0$ where the state is defined. Accordingly, the path integral over the Euclidean counterpart of dS$_2$, namely $\mathbb{S}^2$, computes the inner product $\braket{\Psi|\Psi}$, 
\begin{align}
    \braket{\Psi|\Psi} = \int \mathcal{D}\varphi ~\exp\left(-\int_{-\frac{\pi}{2}}^{\frac{\pi}{2}} L_{\rm dS} ~\sin\theta~d\theta \int_{-\pi}^{\pi} d\phi ~\mathcal{L}[\varphi(\theta,\phi)]   \right) \equiv Z_{\mathbb{S}^2}. 
\end{align}

Furthermore, the correlation functions computed on $\mathbb{S}^2$ can be analytically continued to the Lorentzian signature to get the correlation functions on dS$_2$ evaluated on the no-boundary state via the Schwinger-Keldysh formalism. 
In this way, the Euclidean QFT on $\mathbb{S}^2$ can be thought of as preparing the no-boundary wave function of a state on ${\rm dS}_2$. This interpretation is possible only when the Euclidean manifold on which the QFT is defined admits a time reflection symmetry whose fixed locus separates the manifold into two disconnected parts. In the case above, the $\theta \rightarrow -\theta$ is identified as such a time reflection symmetry, and the $\mathbb{S}^1$ specified by $\theta=0$ is the fixed locus of it.

\begin{figure}[h]      
\centering \includegraphics[width=12cm]{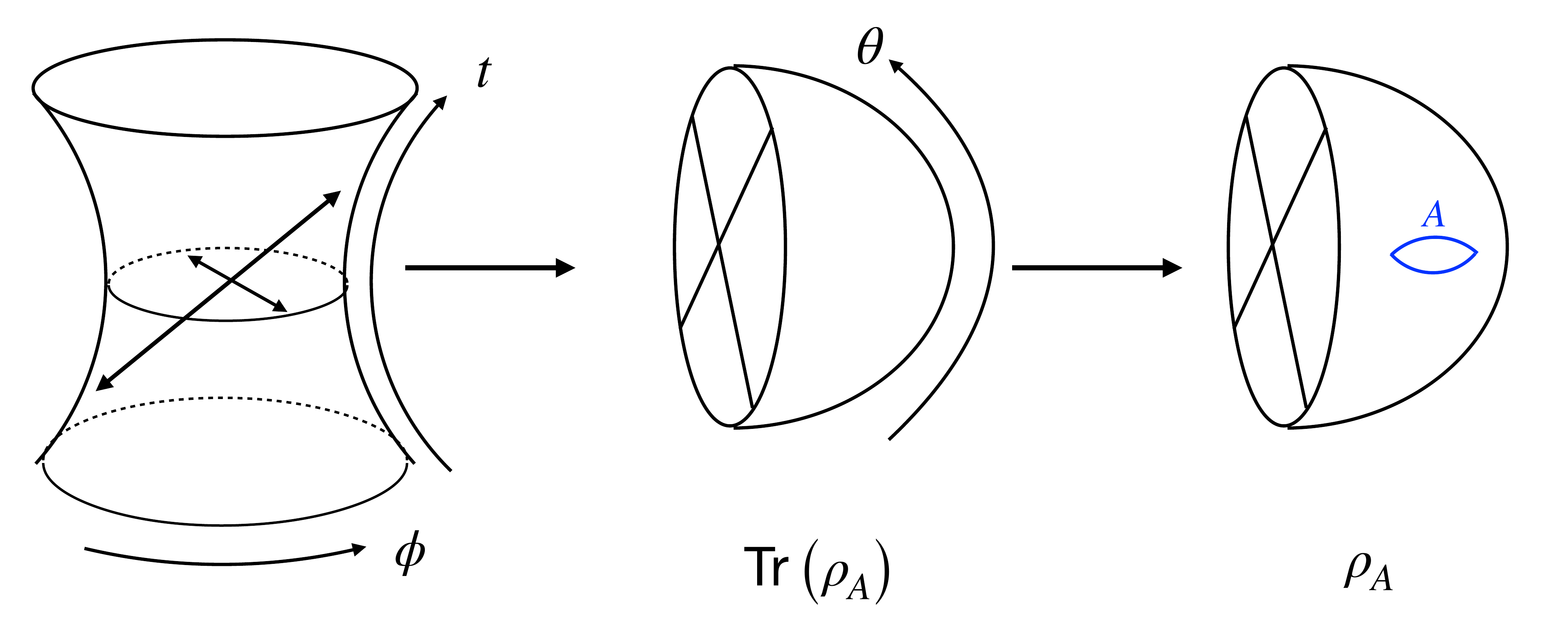}
    \caption{The Euclidean counterpart of dS$_2/\mathbb{Z}_2$ is $\mathbb{RP}^2$, which does not admit a time-reflection symmetry whose fixed locus separates it into two disconnected parts. Therefore, one cannot associate a no-boundary wave function to it. The path integral over the $\mathbb{RP}^2$ with an open slit at $\theta=0$ along the subsystem $A$ computes a Hermitian matrix $\rho_A$. Accordingly, $Z_{\mathbb{RP}^2}={\rm Tr}\rho_A$. We identify this $\rho_A$ as a density matrix on $A$, called the no-boundary density matrix.}
    \label{fig:QFT_RP2} 
\end{figure}

Let us then consider Euclidean QFTs defined on a $\mathbb{RP}^2$, which is obtained by imposing the antipodal identification $(\theta,\phi) \sim (-\theta,\phi+\pi)$ to $\mathbb{S}^2$. Such a $\mathbb{RP}^2$ inherits the time reflection symmetry $\theta \rightarrow -\theta$. The fixed locus of it is the $\mathbb{RP}^1 \cong \mathbb{S}^1$ specified by $\theta = 0$. This fixed locus, however, does not separate the $\mathbb{RP}^2$ into two disconnected parts. 
In fact, not only for the reflection described above, it can be shown that there does not exist any $\mathbb{Z}_2$ isometry of $\mathbb{RP}^2$ such that its fixed locus separates the $\mathbb{RP}^2$ into two disconnected parts. 
Therefore, the $\mathbb{RP}^2$ partition function cannot be interpreted as the inner product of a wave functional under the same logic how the $\mathbb{S}^2$ partition can be. Here, a question arises about how a Euclidean QFT on $\mathbb{RP}^2$ can be linked to a quantum state on ${\rm dS}_2/\mathbb{Z}_2$. 

In order to do so, it is important to notice that a local observer does not have causal access to the whole dS$_2/\mathbb{Z}_2$, although it communicate with almost all the points in it. This means that a local observer can at best sees the density matrix associated to their causal diamond. This motivates us to define the no-boundary density matrix in the following way. 

Consider $\mathbb{RP}^2$, and cut a slit along $\theta = 0$ within the region $A = \{\phi~|~\phi \in [\phi_1,\phi_2]\}$ where $0\leq \phi_1 < \phi_2 < \pi$. This slit induces two boundaries which can be specified by $\theta \rightarrow +0$ and $\theta \rightarrow -0$, respectively. Imposing the boundary condition $\varphi(\pm0, \phi) = \tilde{\varphi}_{\pm}(\phi)$ on the two boundaries and perform the path integral, we obtain a functional with two variables $\tilde{\varphi}_{+}(\phi)$ and $\tilde{\varphi}_{-}(\phi)$, 
\begin{align}\label{eq:rho_A}
    \rho_A[\tilde{\varphi}_{-}(\phi),\tilde{\varphi}_{+}(\phi)] \equiv \int_{\varphi(-0,\phi) = \tilde{\varphi}_-(\phi)}^{\varphi(+0,\phi) = \tilde{\varphi}_+(\phi)}
    \mathcal{D}\varphi ~\exp\left(-\int_{\mathbb{RP}^2} L_{\rm dS} ~\sin\theta~d\theta  d\phi ~\mathcal{L}[\varphi(\theta,\phi)]   \right). 
\end{align}
From this, we can define a matrix on the Hilbert space associated with $A$, 
\begin{align}
    \rho_A \equiv\int \mathcal{D}\tilde{\varphi}_-(\phi) \mathcal{D}\tilde{\varphi}_+(\phi) ~\rho_A[\tilde{\varphi}_{-}(\phi),\tilde{\varphi}_{+}(\phi)]~|\tilde{\varphi}_{-}(\phi)\rangle\langle \tilde{\varphi}_{+}(\phi)|. 
\end{align}
The $\mathbb{RP}^2$ partition function computes ${\rm Tr}(\rho_A)$ accordingly: 
\begin{align}
    {\rm Tr}(\rho_A)  = \int \mathcal{D}\tilde{\varphi}(\phi)  ~\rho_A[\tilde{\varphi}(\phi),\tilde{\varphi}(\phi)] = Z_{\mathbb{RP}^2}. 
\end{align}
If the original Euclidean QFT on $\mathbb{S}^2$ has the time-reflection symmetry, then $\rho_A$ is a Hermitian matrix, thanks to the time-reflection symmetry of $\mathbb{RP}^2$.

We would like to interpret this $\rho_A$ as a (unnormalized) density matrix on $A$ as a subsystem in dS$_2/\mathbb{Z}_2$. We may call such a density matrix the no-boundary density matrix, in the sense that it is prepared by a Euclidean path integral where no extra boundary condition is required except for the two specifying the matrix element. If we perform the same construction starting from a Euclidean QFT on $\mathbb{S}^2$, we will obtain a density matrix on $A$ which can be obtained by performing the partial trace on the no-boundary wave function. Here, on the other hand, we directly define a density matrix from the Euclidean path integral over $\mathbb{RP}^2$, without first introducing a no-boundary wave function.

The philosophy here to define the no-boundary density matrix on the elliptic de Sitter dS$_2/\mathbb{Z}_2$
is very similar to the Ivo-Li-Maldacena (ILM) no-boundary density studied in \cite{ILM24}, but different in the following ways. The ILM no-boundary density matrix is considered in a setup where dynamical gravity is turned on. In their setup, both the geometry and the matter field configuration are fixed on the subregion under the consideration, and all the geometries and field configurations are summed over in a gravitational path integral. It is a generalization of the Hartle-Hawking no-boundary wave function to the density matrix. 
\footnote{It is unclear though whether the ILM no-boundary density matrix obtained in this way can be alternatively obtained by partially tracing out some degrees of freedom in the Hartle-Hawking no-boundary state \cite{HH83}. 
However, by taking this as an assumption, one can impose strong constraints on the ILM no-boundary density matrix as discussed in \cite{BKU25,Wei25}.}.
The no-boundary density matrix we defined in elliptic de Sitter dS$/\mathbb{Z}_2$, on the other hand, is a purely field theoretical object. It is defined on a fixed background and does not involve dynamical gravity. In this case, no no-boundary wave function can be defined on the same background. 

Although we have taken 2D as an example for simplicity, all the arguments apply similarly in higer dimensions. This no-boundary density matrix $\rho_A$ defined on $\mathbb{RP}^2$ is our focus. 

Before proceeding, it is worthwhile to quickly review some conventional interpretations of the $\mathbb{RP}^{d+1}$ partition functions in different contexts, and how the no-boundary density matrix we defined in \eqref{eq:rho_A} is related to them. 
First of all, in the context of 2D CFT, the $\mathbb{RP}^2$ partition function computes the inner product $\braket{C|0}$ where $\ket{0}$ is the ground state in the radial quantization Hilbert space and $\ket{C}$ is a conformal crosscap state \cite{Ishibashi88,FPS93,BP09}.This Hilbert space is defined on the $S^1$ parameterized by $\phi \in (-\pi,\pi]$, and if we regard $A$ on which \eqref{eq:rho_A} is defined as a subsystem of it, $\rho_A$ can be written in terms of a reduced transition matrix \cite{NTTTW20} as $\rho_A={\rm Tr}_{S^1\backslash A} \left(|C\rangle\langle 0|\right)$, which makes the Hermiticity of $\rho_A$ unobvious. This is the stance taken in, e.g. \cite{Wei24}.
This interpretation extends directly to higher dimensions, though they are less studied compared to the 2D case. 
In addition, in the context of dS QFT and gravity, if we use $H_{\rm static}$ to denote the generator of the static patch time evolution, the sphere partition function and the $\mathbb{RP}^2$ partition function can be written as 
\begin{align}
    Z_{\mathbb{S}^2} = {\rm Tr} \left( e^{-2\pi H_{\rm static}} \right),~~~~Z_{\mathbb{RP}^2} = {\rm Tr} \left( e^{-\pi H_{\rm static}} P_{\rm static}\right), 
\end{align}
where $P_{\rm static}$ is the parity transformation in the static patch $\phi \rightarrow \pi-\phi$. See e.g. \cite{CLM11,Law21}.\footnote{Precisely speaking, what these references considered are partition functions on lens spaces in 3D, where $\mathbb{RP}^3$ serves as a special example. Different from even dimensions, $P_{\rm static}$ can be generated by a rotation, and hence the partition function can be interpreted as a grand canonical partition function. 
}   
Similarly, $\rho_A$ can be written as $\rho_A = {\rm Tr}_{{\rm static}\backslash A}(e^{-\pi H_{\rm static}} P_{\rm static})$. It is worth noting that $e^{-\pi H_{\rm static} } P_{\rm static}$ cannot serve as a density matrix since it has negative eigenvalues. However, it is important that $A$ cannot be taken to be the full static patch, not even as a limit. This point will become obvious once we consider the replica method in the next subsection. This relation also extends directly to higher dimensions.

\subsection{Entanglement spectrum, R\'enyi entropy and replica trick}

As a Hermitian operator, the normalized density matrix $\rho_A/{\rm Tr}(\rho_A)$ have real eigenvalues, which is called the entanglement spectrum. The entanglement spectrum is one of the most important information to characterize a density matrix. 

The density matrix can also be characterized by its R\'enyi entropy 
\begin{align}
    S_A^{(n)} \equiv \frac{1}{1-n}
    \log
    \frac{{\rm Tr}\left(\rho_A^n\right)}{\left({\rm Tr}\rho_A\right)^n}~.
\end{align}
If one knows $S_A^{(n)}$ as a function of $n$, one can in principle reconstruct the whole entanglement spectrum $\rho_A/{\rm Tr}(\rho_A)$ from it. 
In the following, we study the R\'enyi entropies of the no-boundary density matrix on the elliptic de Sitter. 
The von Neumann entropy or entanglement entropy
\begin{align}
    S_A \equiv -{\rm Tr} \left(\frac{\rho_A}{\rm Tr\rho_A}\log \frac{\rho_A}{\rm Tr\rho_A}\right) = \lim_{n\rightarrow1} S^{(n)}_A, 
\end{align}
which can be obtained by taking the $n\rightarrow1$ limit of $S_A^{(n)}$, is of particular interest. 

\begin{figure}[h]      
\centering \includegraphics[width=10cm]{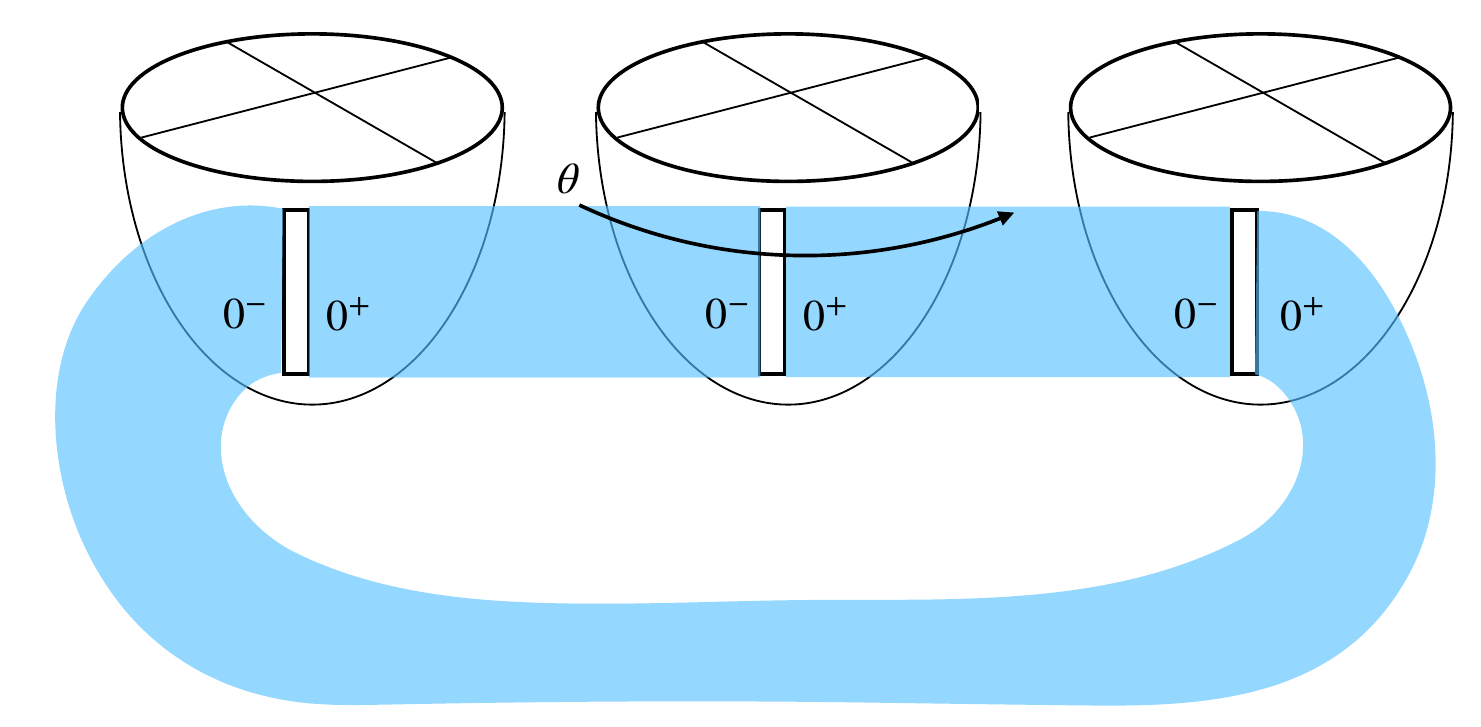}
    \caption{The replica manifold $\Sigma_3$ to compute $\text{Tr}\left(\rho_A^3\right)$. }
    \label{fig:replica Trick} 
\end{figure}

In QFT, the R\'enyi entropy can be computed via the replica trick \cite{CC04} in the following way. Prepare $n$ copies of $\mathbb{RP}^2$ with a slit along $A$ on each of them. Glue the $\theta\rightarrow +0$ boundary in the $i$-th copy to the $\theta\rightarrow -0$ boundary in the $(i+1)$-th copy for $i \in \mathbb{Z}_n$, we obtain a manifold $\Sigma_n$ as shown in figure \ref{fig:replica Trick}. This is called the $n$-replica manifold, and $\Sigma_1 = \mathbb{RP}^2$. In this way, the partition function on $\Sigma_n$ computes ${\rm Tr}(\rho_A^n)$, and the R\'enyi entropy can be obtained as 
\begin{align}
    S_A^{(n)} = \frac{1}{1-n} \log \frac{Z_{\Sigma_n}}{Z_{\Sigma_1}^n}~.
\end{align}
From the next section, we study this in 2D free Dirac fermion CFT as a concrete example.

\section{Replica computation of free fermion on non-orientable surfaces}\label{sec:replica_trick_CFT}

Since $\mathbb{RP}^2$ is a non-orientable surface, we would like to investigate the replica computation in 2D free Dirac fermion CFT on non-orientable surfaces in this section. The results will be used to compute the entanglement entropy and R\'enyi entropies of the no-boundary density matrix on elliptic dS in section \ref{sec:EERP2}. 

We will first explain how the replica trick works for free Dirac fermion CFT in general, then how it is defined on non-orientable surfaces. Finally, we will put them together. The free Dirac fermion CFT serves as an example whose twist operators and conformal crosscap states can be written down explicitly, but the replicated partition function on non-orientable surfaces has not been studied.\footnote{After completing our analysis and during the preparation of this paper, we became aware of two recent papers \cite{Chen25,BTLR25} that also studied replica partition functions in free Dirac fermion CFT on non-orientable surfaces, although they focused on a different physical setup, namely the crosscap quench in 2D CFT \cite{WY24}. We will also show our analysis on this setup in appendix \ref{app:quench}.} Therefore, although the main focus of the current paper is to study the entanglement spectrum of the no-boundary density matrix on dS$/\mathbb{Z}_2$, we expect our analysis provides a starting point of studying entanglement structures of the free Dirac fermion CFT on non-orientable surfaces. 

\subsection{Replica condition and bosonization}

Let us first consider the 2D Dirac fermion CFT defined on a surface parameterized by $(z,\bar{z})$. The free Dirac $(\psi,\bar{\psi})$ fermion can be decomposed into its holomorphic $\psi_L(z)$ (left-moving part) and anti-holomorphic part $\psi_R(\bar{z})$, (right-moving part).

Let us cut a slit along two points $(z_1,\bar{z}_1)$ and $(z_2,\bar{z}_2)$, prepare $n$ copies, and glue them together to create a $n$-replica manifold $\Sigma_n$. A free Dirac fermion $(\psi,\bar{\psi})$ defined on $\Sigma_n$ can be alternatively regarded as $n$ copies of free Dirac fermion $(\psi_j,\bar{\psi}_j)$ ($j=1,2,\cdots,n$) on $\Sigma_1$, which are related by the twist conditions: 
\begin{align}
    &\psi_{L,\,j}(e^{ 2i\pi} (z-z_1))=\psi_{L,\,j+1}( z-z_1)\,,\quad \psi_{L,\,j}(e^{ 2i\pi} (z-z_2))=\psi_{L,\,j-1}(z-z_2)\,, \\
    &\psi_{R,\,j}(e^{ -2i\pi} (\bar{z}-\bar{z}_1))=\psi_{R,\,j+1}(\bar{z}-\bar{z}_1)\,,\quad \psi_{R,\,j}(e^{ -2i\pi} (\bar{z}-\bar{z}_2))=\psi_{R,\,j-1}(\bar{z}-\bar{z}_2)\,,
\end{align}
where only the monodromy matters. 
This can be diagonalized by a discrete Fourier transform of the fermions, 
\begin{equation}
    \psi_L^{(a)}(z)=\frac{1}{\sqrt{n}}\sum_{j} e^{2i\pi \frac{ja}{n}} \psi_{L,\,j}(z)\,,\quad \psi_R^{(a)}(\bar{z})=\frac{1}{\sqrt{n}}\sum_{j} e^{-2i\pi \frac{ja}{n}} \psi_{R,\,j}(\bar{z})\,.
\end{equation}
This procedure diagonalizes the theory, reducing the computation to $n$ independent copies with no interactions between them, aside from the insertion of a topological operator in each copy. The boundary conditions involve only a single copy of the CFT:
\begin{equation}
     \psi_L^{(a)}(e^{2i\pi}(z-z_1))=e^{2 i\pi a}\psi_L^{(a)}(z-z_1)\,,\quad \psi_L^{(a)}(e^{i2\pi}(z-z_2))=e^{-2 i\pi a}\psi_L^{(a)}(z-z_2)\,.
     \label{eq:twisted boundary condtions}
\end{equation}
This can be engineered by adding two twist operators at positions $(z_{1},\bar{z}_{1})$ and$(z_{2},\bar{z}_{2})$ on each copy of the CFT and computing two-point functions in this background. The final result is therefore the product of the two-point functions on each copy:
\begin{equation}
    Z_{\Sigma_n} = \prod_{a=-\frac{n-1}{2}}^{\frac{(n-1)}{2}}\left<\sigma^{(a)}(z_{1},\bar{z}_1)\sigma^{(-a)}(z_{2},\bar{z}_{2})\right>\,.
    \label{Rényi}
\end{equation}
where $\sigma^{(a)}$ denotes the twist operators, which can be written down explicitly via the bosonization of the Dirac fermion fields
\begin{equation}
    \psi_L(z)=e^{i X_L(z)}\,,\quad \bar{\psi}_L(z)=e^{-iX_L(z)}\,,\quad \psi_R(\bar{z})=e^{i X_R(\bar{z})}\,, \quad \bar{\psi}_R(\bar{z})=e^{-i X_R(\bar{z})}\,,
\end{equation}
where $X_L(z)$ ($X_R(\bar{z})$) is the holomorphic (anti-holomorphic) part of the free boson field. 
It turns out there are two different classes of twist operators which can be constructed to implement \eqref{eq:twisted boundary condtions} as \cite{TU10}
\begin{equation}\label{eq:twist_operator}
    \sigma^{(a,\,\pm)}(z,\bar{z})= e^{i\frac{a}{n}\left(X_L(z)\pm X_R(\bar{z})\right)}\,.
\end{equation}
It was shown in \cite{OT14} that twist operators can be interpreted as conformal boundary conditions for the fermions, in particular $\sigma^{(a,\,+)}$ corresponds to Dirichlet boundary conditions while $\sigma^{(a,\,-)}$ corresponds to Neumann boundary conditions. We will shortly argue that there is an analogous sign choice in the construction of the crosscaps, and that there is a compatibility condition between this choice of sign, and the one of the twist operator $\sigma^{a,\,\pm}$ when combining them together.

\subsection{CFT on non-orientable surfaces and crosscap states}
Any non-orientable surface can be constructed by starting from an orientable surface and then inserting crosscaps on it. Therefore, we call CFT defined on non-orientable surfaces the crosscap CFT (XCFT). In string worldsheet theory for propagating unoriented strings, summing over both orientable and non-orientable string worldsheets is considered \cite{Polchinski98-1,Polchinski98-2}. This might be the first encounter of XCFT for most readers. Recently, XCFT and its lattice counterparts have attracted attentions from multiple different perspectives. For example, it can be used to extract universal data in critical lattice systems \cite{Tu17,ZHZTWT22,SU24}, efficiently prepare thermal pure states \cite{CY24,Yoneta24,CK21}, engineer new quench dynamics where scrambling of correlations can be seen within thermal equilibrium \cite{CCR24,WY24}, and embed dS$_2$ structures into AdS$_3$/CFT$_2$ \cite{Wei24}. New types of crosscap conditions have also been recently constructed in both lattice systems \cite{ZWWT24} and CFTs \cite{HKKL25}. 

Let us start from a Riemann sphere and use the standard holomorphic coordinate $z$ to parameterize it. The simplest non-orientable surface is $\mathbb{RP}^2$ which can be obtained by the following $\mathbb{Z}_2$ quotient of the Riemann sphere:
\begin{equation}
    z \sim -\frac{1}{\bar{z}} \,.
\end{equation}
The $z\bar{z}\leq 1$ region serves as a fundamental domain of $\mathbb{RP}^2$, where antipodal points on the ``boundary" $z\bar{z} = 1$ are identified with each other. 

Let us perform a Weyl transformation to make the $z\bar{z}< 1$ a unit disk with a flat metric. The partition function can be then regarded as the transition amplitude $\braket{C|0}$ between the CFT ground state $\ket{0}$ and a conformal crosscap state $\ket{C}$ living in the Hilbert space of the radial quantization. Because of the antipodal identification on $z\bar{z} = 1$, the conservation of energy reads:
\begin{equation}
    T(z)=\frac{\bar{T}\left(\bar{z}=-\frac{1}{z}\right)}{z^4}\,.
\end{equation}
This condition imposes a restriction on $\ket{C}$ \cite{Ishibashi88,Cardy89,BP09}, 
\begin{equation}
\label{eq:Ishibashi states +}
    L_{n}\ket{C}=(-1)^n\bar{L}_{-n}\ket{C},
\end{equation} 
where $L_n$'s ($\bar{L}_n$'s) are the holomorphic (anti-holomorphic) Virasoro generators. 

In free Dirac CFT, the conformal crosscap states can be written down explicitly after bosonization. Decomposing the free boson field $X(z,\bar{z})$ into
\begin{align}
    X(z,\bar{z})&=X_L(z)+X_R(\bar z), 
\end{align}
we have 
\begin{align}
    X_L(z)&=x_L-i\frac{ \alpha'}{2} p_L \log(z)+i \frac{\alpha'}{2} \sum_{k \in \mathbb{Z}*}\frac{\alpha_k}{k ~z^k}\\
    X_R(\bar{z})&=x_R-i \frac{ \alpha'}{2} p_R\log(\bar{z})+i \frac{\alpha'}{2} \sum_{k \in \mathbb{Z}*}\frac{\tilde{\alpha}_k}{k ~\bar{z}^k}\,.
\end{align}
{with $p_L=\frac{m}{R}+\frac{w R}{\alpha'}$ and $p_R=\frac{m}{R}-\frac{w R}{\alpha'}$ , $m$ the momentum and $w$ the winding number. $R$ is the radius of the compact circle, $\alpha'$ the string tension and $(x_L,x_R)$ are the center of mass positions.} The commutation relations for the oscillatory part follow:
\begin{equation}
    [x_L,p_L]=i\,,\quad[x_R,p_R]=i\,,\quad [\alpha_k,\alpha_q]=k \delta_{k+q,0}\,,\quad[\tilde{\alpha}_k,\tilde{\alpha}_q]=k \delta_{k+q,0}\,.
\end{equation}
and the identification: $\alpha_0=p_L,\,\tilde{\alpha}_0=p_R$. The condition \eqref{eq:Ishibashi states +} has two possible implementations:
\begin{equation}
   \alpha_{k}\ket{C_+}=(-1)^k \tilde{\alpha}_{-k}\ket{C_+}\,\quad \text{or}\,\quad 
    \alpha_{k}\ket{{C}_-}=-(-1)^k \tilde{\alpha}_{-k}\ket{{C}_-}\,.
    \label{eq:crosscap condition} 
\end{equation}
The vacuum of the free boson is characterized by its zero modes: center of mass momentum $m$ and winding $w$. The condition on the zero mode for $\ket{{C}_+}$ imposes that winding modes should vanish, while momentum should vanish for ${C}_-$. As a result, we obtain the conformal crosscap states:
\begin{align}
    \ket{{C}_+}&=\exp\left(~~~\sum_{k=1}^{\infty} \frac{(-1)^k}{k}\alpha_{-k}\tilde{\alpha}_{-k}\right)\sum_{m}C_{m}\ket{m,0}\\
    \ket{{C}_-}&=\exp\left(- \sum_{k=1}^{\infty} \frac{(-1)^k}{k}\alpha_{-k}\tilde{\alpha}_{-k}\right)\sum_{w}C_{w}\ket{0,w}\,.
\end{align}
with $\ket{m,w}$ the zero mode sector with $\alpha_0\ket{m,w}=p_L\ket{m,w}$ and $\tilde{\alpha}_0\ket{m,w}=p_R\ket{m,w}$. It has been shown in \cite{Tan:2024dcd} that the $C_m$ and $C_w$ coefficient are given by:
\begin{equation}\label{eq:crosscap_states}
    C_{m}=\begin{cases}
0 & \text{$m$ is odd} \\
1  & \text{$m$ is even}
\end{cases}\qquad \qquad C_{w}=\begin{cases}
0 & \text{$w$ is odd} \\
1  & \text{$w$ is even .}
\end{cases}
\end{equation}

\subsection{Example: replica computation on \texorpdfstring{$\mathbb{K}^2$}{K2}}\label{sec:K2}

With the explicit expressions of the twist operators \eqref{eq:twist_operator} and conformal crosscap states \eqref{eq:crosscap_states}, now we are ready to perform the replica computation on non-orientable surfaces. 

As an explicit example, let us consider a finite cylinder whose $\mathbb{S}^1$ direction is parameterized by $l \in [0,2\pi)$ with  and interval direction parameterized by $\tau\in[0,2\alpha]$. By identifying the antipodal points on each boundary, we obtain a Klein bottle $\mathbb{K}^2$, whose partition function can be written as 
\begin{align}
    Z_{\mathbb{K}^2} = \braket{C|e^{-2\alpha H}|C}, 
\end{align}
where $H$ is the Hamiltonian associated to the radial quantization. Let us use 
\begin{align}
    (y,\bar{y}) \equiv (\tau + il, \tau-il )
\end{align}
to parameterize it. Let us cut a slit between $(y_1,\bar{y}_1)$ and $(y_2,\bar{y}_2)$, and consider the $n$ replica of it. We will need to compute following two point functions of the twist operators: 
\begin{align}
    \bra{C}e^{-2\alpha H}\sigma^{(a)}(y_1,\bar{y}_1)\sigma^{(-a)}(y_2,\bar{y}_2)\ket{{C}}. 
\end{align}
At this point there are several choices of sign which can be considered: $\sigma^{(a,\,\pm)}$ and $\ket{{C}_\pm}$. We explain in the appendix \ref{sec:appendix two-point function KB} that we should choose the combinations $\left(\sigma^{(a,\,-)},~\ket{{C}_-}\right)$ and $\left(\sigma^{(a,\,+)},~\ket{{C}_+}\right)$. With e.g. the $\left(\sigma^{(a,\,-)},~\ket{{C}_-}\right)$ combination, we obtain two point function
\begin{align}
    &\frac{\bra{{C}}e^{-2\alpha H}\sigma^{(a)}(y_1,\bar{y}_1)\sigma^{(-a)}(y_2,\bar{y}_2)\ket{{C}}}{\bra{{C}}e^{-2\alpha H}\ket{{C}}}=\frac{\theta_3(\frac{a}{n}\frac{1}{4\pi i}\left(y_1-y_2+\bar{y}_1-\bar{y}_2\right)|\frac{2 i \alpha}{\pi})}{\theta_3(0|\frac{2 i \alpha}{\pi})}\nonumber \\
    &\times\left(\frac{\eta\left(\frac{2i\alpha}{\pi}\right)^6\theta_1\left(\frac{y_1+\bar{y}_2}{2\pi i}+\frac{i \pi}{2\pi i}\mid\frac{2i\alpha}{\pi}\right)\theta_1\left(\frac{y_2+\bar{y}_1}{2\pi i}+\frac{i \pi}{2\pi i}\mid\frac{2i\alpha}{\pi}\right)}{\theta_1\left(\frac{y_2+\bar{y}_2}{2\pi i}+\frac{i \pi}{2\pi i}\mid\frac{2i\alpha}{\pi}\right)\theta_1\left(\frac{y_1+\bar{y}_1}{2\pi i}+\frac{i \pi}{2\pi i}\mid\frac{2i\alpha}{\pi}\right)\theta_1\left(\frac{y_2-y_1}{2\pi i}\mid\frac{2i\alpha}{\pi}\right)\theta_1\left(\frac{\bar{y}_2-\bar{y}_1}{2\pi i}\mid\frac{2i\alpha}{\pi}\right)}\right)^{\left(\frac{a}{ n}\right)^2}.
\end{align}
Here we have set $\alpha'=2$ and $R=1$. 
Assuming $\tau_1 = \tau_2 = \tau$, i.e. 
\begin{equation}
\left(y_1,\bar{y}_1\right)=\left(\tau+il_1,\tau-il_1\right)\,,\quad \left(y_2,\bar{y}_2\right)=\left(\tau+il_2,\tau-il_2\right)\,,
\end{equation}
From the two-point function we can compute the R\'eyni entropies: 
\begin{align}\label{eq:twist_2p}
    \frac{\mathrm{Tr}\left(\rho_A^n\right)}{\left(\mathrm{Tr}\rho_A\right)^n}=&\prod_{a=-\frac{n-1}{2}}^{\frac{(n-1)}{2}}\frac{\bra{{C}}e^{-2\alpha H}\sigma^{(a)}(y_1,\bar{y}_1)\sigma^{(-a)}(y_2,\bar{y}_2)\ket{{C}}}{\bra{{C}}e^{-2\alpha H}\ket{{C}}}\\
    &=\left(\frac{\eta\left(\frac{2i\alpha}{\pi}\right)^6\theta_1\left(\frac{y_1+\bar{y}_2}{2\pi i}+\frac{i \pi}{2\pi i}\mid\frac{2i\alpha}{\pi}\right)\theta_1\left(\frac{y_2+\bar{y}_1}{2\pi i}+\frac{i \pi}{2\pi i}\mid\frac{2i\alpha}{\pi}\right)}{\theta_1\left(\frac{y_2+\bar{y}_2}{2\pi i}+\frac{i \pi}{2\pi i}\mid\frac{2i\alpha}{\pi}\right)\theta_1\left(\frac{y_1+\bar{y}_1}{2\pi i}+\frac{i \pi}{2\pi i}\mid\frac{2i\alpha}{\pi}\right)\theta_1\left(\frac{y_2-y_1}{2\pi i}\mid\frac{2i\alpha}{\pi}\right)\theta_1\left(\frac{\bar{y}_2-\bar{y}_1}{2\pi i}\mid\frac{2i\alpha}{\pi}\right)}\right)^{\left(n-\frac{1}{n}\right)}\,, \nonumber
\end{align}
where we have used $\sum_{a=-\frac{n-1}{2}}^{\frac{n-1}{2}}\left(\frac{a}{n}\right)^2=\frac{1}{12}\left(n-\frac{1}{n}\right)$.

This example can be related to the entanglement entropy in the crosscap quench \cite{WY24} by 
taking the limit $n \rightarrow 1$ of $\frac{1}{n-1}\mathrm{Tr}\left(\rho_A^n\right)$, performing the continuation $\tau \rightarrow \alpha + it$, and introducing a new parameter $\beta = 4\alpha$: 
\begin{align}
    S_A(t,&\,l) =
    &\frac{1}{6} \log \left(\frac{1}{\varepsilon_{\rm UV}^2}\frac{\left|\theta_1 \left(\frac{l}{2\pi}|\frac{i\beta}{2\pi}\right)\right|^2 \cdot \left|\theta_1 \left(\frac{1}{\pi}\left(t+\frac{\pi}{2}-i\frac{\beta}{4}\right)|\frac{i\beta}{2\pi}\right)\right|^2}
    {\eta\left(\frac{i\beta}{2\pi}\right)^6 \cdot \left|\theta_1 \left(\frac{1}{\pi}\left(\frac{l}{2}+t+\frac{\pi}{2}-i\frac{\beta}{4}\right)|\frac{i\beta}{2\pi}\right)\right| 
    \cdot
    \left|\theta_1 \left(\frac{1}{\pi}\left(-\frac{l}{2}+t+\frac{\pi}{2}-i\frac{\beta}{4}\right)|\frac{i\beta}{2\pi}\right)\right|}\right),  \nonumber
\end{align}
which we will discuss in appendix \ref{app:quench}.


\section{Entanglement entropy in elliptic de Sitter}
\label{sec:EERP2}

Based on the tools developed in the last section, we compute the entanglement entropy and the R\'enyi entropy of the no-boundary density matrix in dS$_2/\mathbb{Z}_2$ and its real time evolution in this section. Since we have already computed the 2-point function of twist operators on $\mathbb{K}^2$ in section \ref{sec:K2}, let us utilize the results there to obtain the two-point function of twist operators on the $\mathbb{RP}^2$ as follows. In this section, we set the dS radius $L_{\rm dS} = 1$. 

Consider again the $\mathbb{K}^2$ described in section \ref{sec:K2}. Taking $\alpha \rightarrow \infty$, we obtain a half cylinder with one crosscap at $\tau = 0$. The two-point function of twist operators reads
\begin{align}
    \left<\sigma^{(a)}(\tau_1,l_1)\sigma^{(-a)}(\tau_2,l_2)\right>_{\rm half ~cylinder}\equiv &\lim_{\alpha\rightarrow \infty}\frac{\bra{C}e^{-2\alpha H}\sigma^{(a)}(\tau_1,l_1)\sigma^{(-a)}(\tau_2,l_2)\ket{C}}{\bra{C}e^{-2\alpha H}\ket{C}}\!\! \nonumber\\=&\left(\!\frac{\!\left|\sinh\!\left(\frac{\tau_1+\tau_2}{2}+i\frac{\pi+(l_1-l_2)}{2}\right)\sinh\left(\!\frac{\tau_1+\tau_2}{2}+i\frac{\pi-(l_1-l_2)}{2}\right)\!\right|}{|\sin(\frac{l_1-l_2}{2})|^2|\sinh(\tau_1+i\frac{\pi}{2})||\sinh(\tau_2+i\frac{\pi}{2})|}\!\!\right)^{\!\!\left(\frac{a}{ n}\right)^2}, 
\end{align}
where 
\begin{equation}
\left(y_1,\bar{y}_1\right)=\left(\tau_1+il_1,\tau_1-il_1\right)\,,\quad \left(y_2,\bar{y}_2\right)=\left(\tau_2+il_2,\tau_2-il_2\right)\,.
\end{equation}
and $\alpha' =2$. Let us then perform the conformal transformation 
\begin{align}
    (z=e^{y},\bar{z}=e^{\bar{y}}),
\end{align}
under which the half cylinder is mapped to the $z\bar{z}\geq1$ region of a complex plane with flat metric 
\begin{align}
    ds^2 = dzd\bar{z}, 
\end{align}
with points on $z\bar{z}=1$ antipodally identified with each other. Let us call this configuration the flat $\mathbb{RP}^2$. Since $\sigma^{(a)}$ is a conformal primary of weight $\left((\frac{a}{n})^2,(\frac{a}{n})^2\right)$, we obtain
\begin{align}
    &\left<\sigma^{(a)}(z_1,\bar{z}_1)\sigma^{(-a)}(z_2,\bar{z}_2)\right>_{\rm flat~\mathbb{RP}^2} \nonumber\\
    =&\left(\frac{{\rm d}z_1}{{\rm d}y_1}\right)^{-\left(\frac{a}{n}\right)^2}\left(\frac{{\rm d}\bar{z}_1}{{\rm d}\bar{y}_1}\right)^{-\left(\frac{a}{n}\right)^2}\left(\frac{{\rm d}z_2}{{\rm d}y_2}\right)^{-\left(\frac{a}{n}\right)^2}\left(\frac{{\rm d}\bar{z}_2}{{\rm d}\bar{y}_2}\right)^{-\left(\frac{a}{n}\right)^2}\left<\sigma^{(a)}(y_1,\bar{y}_1)\sigma^{(-a)}(y_2,\bar{y}_2)\right>_{\rm half ~cylinder}\nonumber\\ =&\left(\frac{|z_2\bar{z}_1+1||z_1\bar{z}_2+1|}{|z_1-z_2|^2|z_1\bar{z}_1+1||z_2\bar{z}_2+1|}\right)^{\left(\frac{a}{n}\right)^2}\,.
    \label{eq:two pt function flat RP2}
\end{align}
As the next step, we perform the Weyl transformation to the metric such that it goes back to the spherical metric \eqref{eq:RP2metric},
\begin{align}
    ds^2= d\theta^2 + \cos^2 \theta d~\phi^2 =\frac{4}{(1+z \bar{z})^2} dzd\bar{z} \equiv e^{2\omega(z,\bar{z})} dzd\bar{z}, 
\end{align}
where we have set $L_{\rm dS} = 1$ for simplicity and 
\begin{align}
    (z,\bar{z}) = (e^{i\phi} \tan\big((\pi/2 - \theta)/2\big), e^{-i\phi} \tan\big((\pi/2 - \theta)/2\big)). 
\end{align}
The two-point function then becomes 
\begin{align}
    \left<\sigma^{(a)}(z_1,\bar{z}_1)\sigma^{(-a)}(z_2,\bar{z}_2)\right>_{\mathbb{RP}^2} =& \left(e^{- \omega(z_1,\bar{z}_1})\right)^{\left(\frac{a}{n}\right)^2} \left(e^{- \omega(z_2,\bar{z}_2)}\right)^{\left(\frac{a}{n}\right)^2}\left<\sigma^{(a)}(z_1,\bar{z}_1)\sigma^{(-a)}(z_2,\bar{z}_2)\right>_{\rm flat~\mathbb{RP}^2}\nonumber\\
    =&\left(\frac{1}{4}\frac{|z_2\bar{z}_1+1||z_1\bar{z}_2+1|}{|z_1-z_2|^2}\right)^{\left(\frac{a}{n}\right)^2}\,,
\end{align}
and with the $(\theta, \phi)$ coordinates: 
\begin{align}
    &\left<\sigma^{(a)}(\theta_1,\phi_1)\sigma^{(-a)}(\theta_2,\phi_2)\right>_{\mathbb{RP}^2} \nonumber\\
    =&\left(\frac{\left|e^{i(\phi_2-\phi_1)}\tan\left(\frac{\frac{\pi}{2}-\theta_1}{2}\right)\tan\left(\frac{\frac{\pi}{2}-\theta_2}{2}\right)+1\right|\left|e^{i(\phi_1-\phi_2)}\tan\left(\frac{\frac{\pi}{2}-\theta_1}{2}\right)\tan\left(\frac{\frac{\pi}{2}-\theta_2}{2}\right)+1\right|}{4~\left|e^{i(\phi_1-\phi_2)}\tan\left(\frac{\frac{\pi}{2}-\theta_1}{2}\right)-\tan\left(\frac{\frac{\pi}{2}-\theta_2}{2}\right)\right|^2}\right)^{\left(\frac{a}{n}\right)^2}.\nonumber
\end{align}
{\red


}
By construction, $|z|\ge 1$ corresponds to $\theta \in [-\frac{\pi}{2}, 0]$, with a crosscap at $\theta=0$. The choice of fundamental domain can, however, be changed freely. We take it to be $\theta \in [-\frac{\pi}{2}, \frac{\pi}{2}]$ and $\phi \in [0, \pi]$ as shown in figure \ref{fig:QFT_RP2} so that the time reflection symmetry under $\theta \rightarrow -\theta$ manifests.

The entanglement entropy and R\'enyi entropies for a subsystem $A$ extends between $(\theta_1,\phi_1)$ and $(\theta_2,\phi_2)$ read
\begin{align}\label{eq:EE_RP2}
    S_A 
    = \frac{1}{6}\log\left(\frac{4~\left|e^{i(\phi_1-\phi_2)}\tan\left(\frac{\frac{\pi}{2}-\theta_1}{2}\right)-\tan\left(\frac{\frac{\pi}{2}-\theta_2}{2}\right)\right|\left|e^{i(\phi_1-\phi_2)}\tan\left(\frac{\frac{\pi}{2}+\theta_1}{2}\right)-\tan\left(\frac{\frac{\pi}{2}+\theta_2}{2}\right)\right|}{\varepsilon_{\rm UV}^2\left|e^{i(\phi_2-\phi_1)}\tan\left(\frac{\frac{\pi}{2}-\theta_1}{2}\right)\tan\left(\frac{\frac{\pi}{2}-\theta_2}{2}\right)+1\right|\left|e^{i(\phi_2-\phi_1)}\tan\left(\frac{\frac{\pi}{2}+\theta_1}{2}\right)\tan\left(\frac{\frac{\pi}{2}+\theta_2}{2}\right)+1\right|}\right).
\end{align}
and 
\begin{align}
    S^{(n)}_A = \frac{1}{2}\left(1+\frac{1}{n}\right) S_A. 
\end{align}
where $\varepsilon_{\rm UV}$ is a UV cutoff corresponding to the minimal length unit in the expanding/shrinking universe \cite{CS22}.

By continuing the imaginary time to real time $\theta_{1(2)} \rightarrow it_{1(2)}$, we can get the entanglement entropy for subsystems in dS$_2/\mathbb{Z}_2$. Let us study \eqref{eq:EE_RP2} in more details in some special cases. 

\paragraph{No-boundary density matrices at $t=0$}~\par
Let us start by looking at the no-boundary density matrices living on the equator, i.e. the initial time slice $\t=0$. In this case,  
\begin{align}
    S_A|_{t=0} = \frac{1}{3}\log \left(\frac{2}{\varepsilon_{\rm UV}} \tan\frac{|\phi_2-\phi_1|}{2}\right). 
\end{align}
For a subsystem which is small enough $|\phi_2-\phi_1|\ll1$, 
\begin{align}
    S_A|_{t=0} \approx \frac{1}{3} \log \frac{|\phi_2 - \phi_1|}{\varepsilon_{\rm UV}},~~(|\phi_2-\phi_1|\ll1) ,
\end{align}
reproducing the entanglement entropy formula in an infinitely large system \cite{CC04}. On the other hand, if we look at the limit $\pi-|\phi_2-\phi_1|\ll1$ where the subsystem $A$ approaches the whole spatial region, 
\begin{align}
    S_A|_{t=0} \approx \frac{1}{3} \log \frac{4}{\varepsilon_{\rm UV}|\phi_2-\phi_1|},~~(\pi-|\phi_2-\phi_1|\ll1) .
\end{align}
This is a stronger divergence than that appears in the entanglement entropy in global dS spacetime\cite{MP12} when the subsystem approaches the half of the spatial region. This also indicates that the state living on the whole spatial region, if any, should not be a pure state. This is consistent with the fact that we cannot define a no-boundary wave function via the Euclidean path integral on $\mathbb{RP}^2$ discussed in section \ref{sec:EdS_NBDM} and that the standard canonical quantization collapses on the global spatial slice discussed in section \ref{sec:1D_space}. 

\paragraph{$S_A$ off the equator $\theta\neq 0$}~\par
\begin{figure}[h]    
    \centering
    \includegraphics[width=0.5\linewidth]{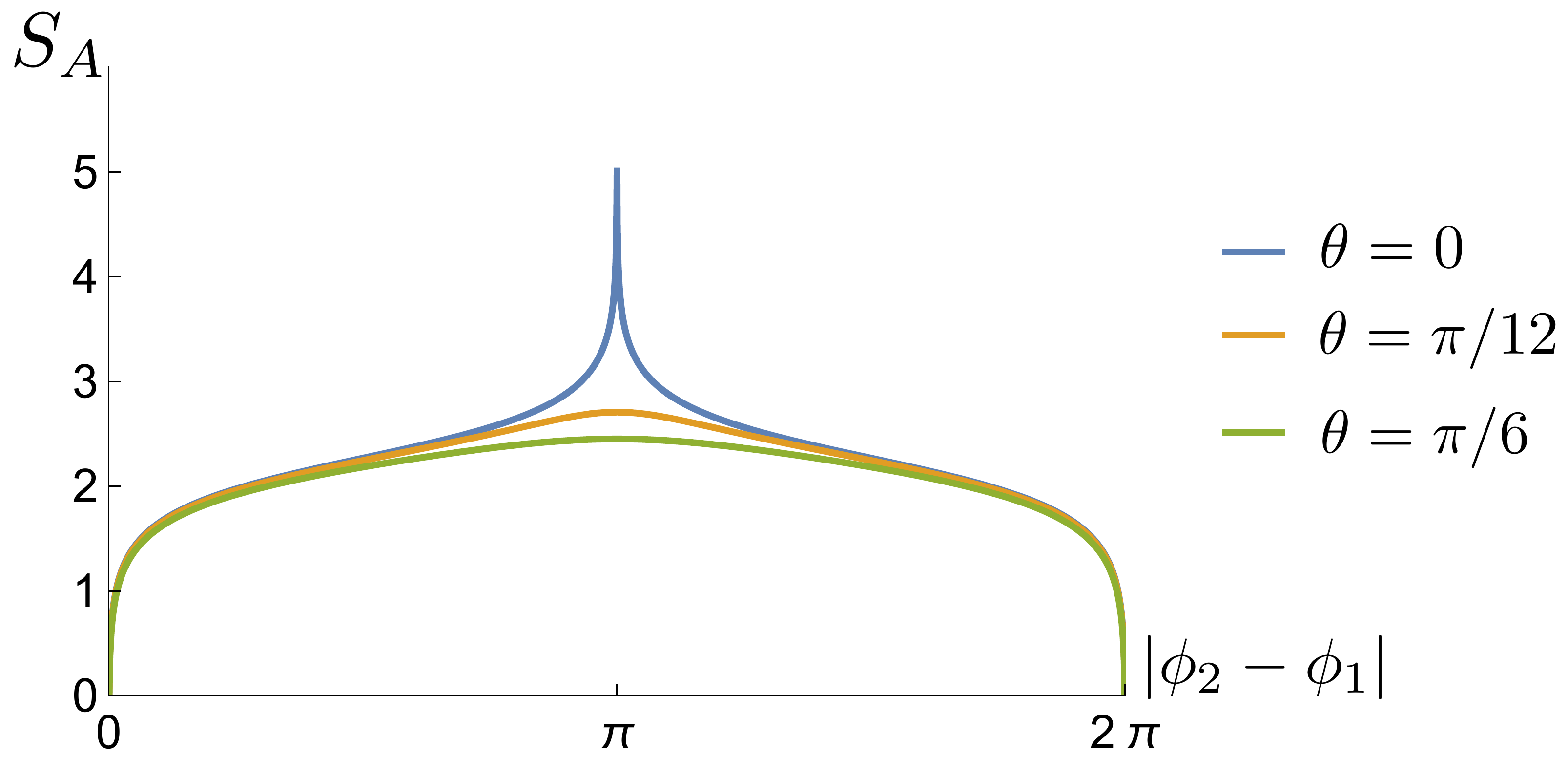}
    \caption{The $S_A$ for single-interval subsystem $A$ is shown as a function of $|\phi_2-\phi_1|$ for different Euclidean times $\theta_1=\theta_2=\theta$. Here we have set $\varepsilon_{\rm UV}=4\exp(-15/2)\approx 0.002$}
    \label{fig:entanglement entropy fix theta function phi}
\end{figure}

 
Let us comment on the behavior of the entanglement entropy formula \eqref{eq:EE_RP2} off the equator by setting $\theta_1=\theta_2=\theta \neq 0$. In this case, the corresponding replica computation lacks a time-reflection symmetry and cannot be actually interpreted as the entanglement entropy for a density matrix anymore, but the pseudo entropy for a transition matrix instead \cite{NTTTW20,MSTTW20}. However, it is still useful to know its behavior. In this case, \eqref{eq:EE_RP2} is simplified to 
\begin{align}\label{eq:equal_theta}
    S_A
    =\frac{1}{6}\log\left(\frac{16\sin^2(\frac{\phi_1-\phi_2}{2}) \cos^2(\theta)}{\varepsilon_{\rm UV}^2\left(3+2 \cos(\phi_1-\phi_2)\cos^2(\theta)-\cos(2\theta)\right)}\right).
\end{align}
Figure \ref{fig:entanglement entropy fix theta function phi} shows the $|\phi_2-\phi_1|$ dependence for $S_A$ with fixed $\theta$'s. 

\paragraph{Real time evolution of entanglement entropy: fixed proper length cases}~\par
Let us then consider an interval $A$ with a fixed proper length in dS$_2/\mathbb{Z}_2$ evolving in time. This serves as a nontrivial check of our formula \eqref{eq:EE_RP2}, since the entanglement entropy should not depend on the time but only on the proper length due to the symmetry of dS$_2/\mathbb{Z}_2$, yet the computation is nontrivial. 

Let the two endpoints of $A$ to be 
\begin{align}
    (t_1,\phi_1) = (t,-\phi(t)),~~~(t_2,\phi_2) = (t,+\phi(t)),
\end{align}
where $\phi(t) \in [0,\frac{\pi}{2}]$. The geodesic equation between a pair of spacelike separated points can be parametrized by:
\begin{align}
    -\left(\frac{dt}{d\lambda}\right)^2+ \cosh(t)^2 \left(\frac{d\phi}{d\lambda}\right)^2=1,~~~ ~~~\cosh(t)^2~\frac{d\phi}{d\lambda}=E\,.
\end{align}
where $\lambda$ denotes the affine parameter and we have used that $\partial_{\phi}$ is a killing vector leading to the conserved quantity $E$. The parameter $E$ can be related to $E=\cosh(t_0)$, where $t_0$ is the maximal time reached by a geodesic connecting $(t, \phi(t))$ and $(t, -\phi(t))$. It is then straightforward to obtain:
\begin{align}
    2l =\int d\lambda=2\int_{t}^{t_0} \frac{dt'}{\sqrt{\left(\frac{\cosh(t_0)}{\cosh(t')}\right)^2-1}},~~~\phi=\int d\phi=\int_{t}^{t_0} \frac{dt'}{\cosh(t')\sqrt{\cosh(t_0)^2-\cosh(t')^2} }\,. \nonumber
\end{align}
From which we deduce:
\begin{align}
    \tan\left(\frac{\pi}{2}-l\right)=\sqrt{2}\frac{\sinh(t)}{\sqrt{\cosh(2 t_0)-\cosh(2t)}},~~~\tan\left(\frac{\pi}{2}-\phi\right)=\frac{\sqrt{1+\cosh(2 t_0)}\sinh(t)}{\sqrt{\cosh(2 t_0)-\cosh(2t)}}\,.
\end{align}
This system can be inverted to obtain $\phi(t)$ as a function of the physical parameters $(t,l)$, where $2l$ is the proper length of the subsystem, it leads to:
\begin{equation}
    \tan(\phi(t,l))=\tan(l) \sqrt{\frac{1+\cos(2l)}{\cos(2l)+\cosh(2t)}}\,.
    \label{eq:phi as a function of proper length}
\end{equation}
Substituting \eqref{eq:phi as a function of proper length} to \eqref{eq:EE_RP2}, we obtain 
\begin{align}
    S_A|_{{\rm proper~length}=2l} = \frac{1}{3}\log \left(\frac{2}{\varepsilon_{\rm UV}} \tan l\right), 
\end{align}
which indeed does not depend on $t$ as expected.

\paragraph{Real time evolution of entanglement entropy: co-expanding cases}~\par

\begin{figure}[h]
    \centering
    \includegraphics[width=0.6\linewidth]{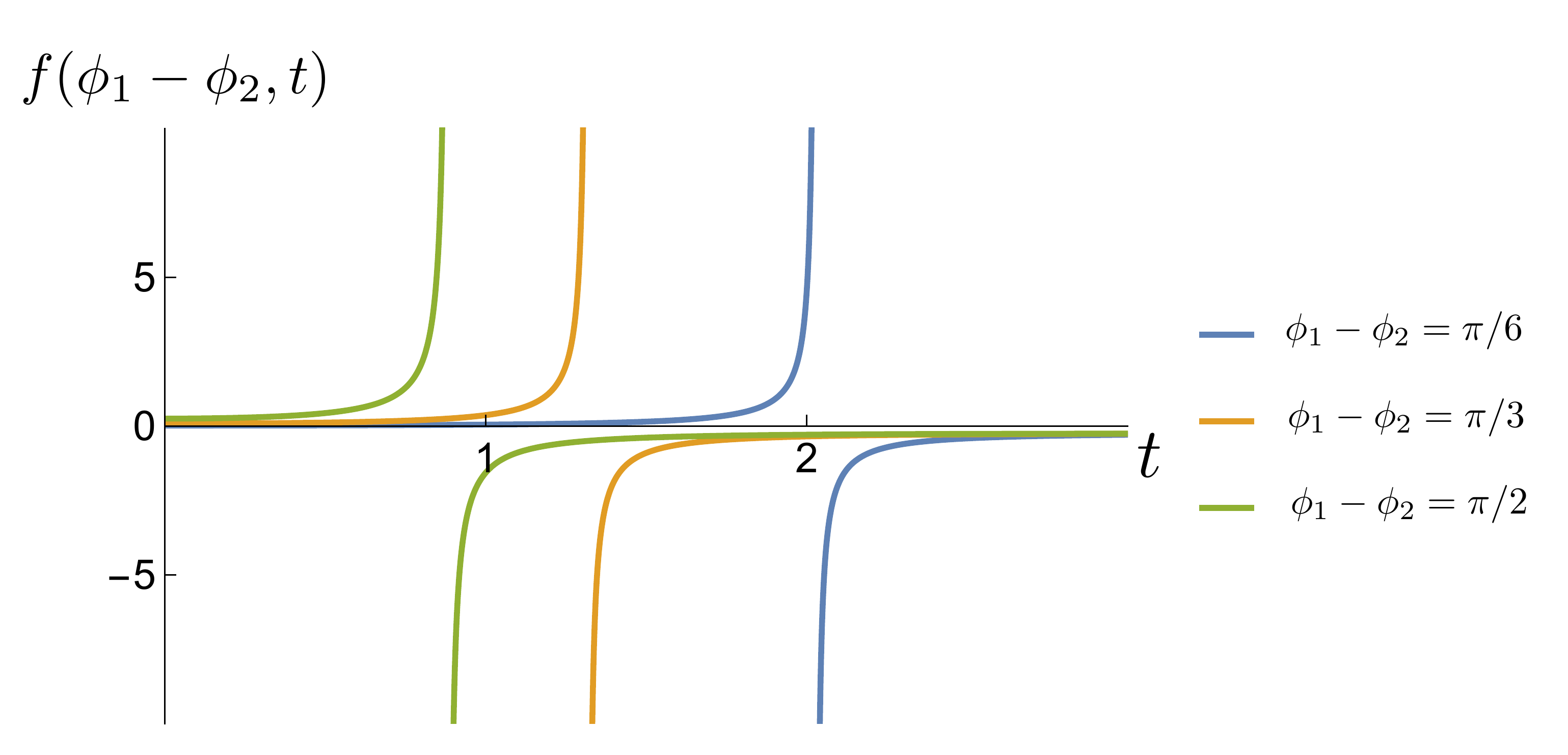}
    \caption{$f(\phi_1-\phi_2,t)$ as a function of $t$ for fixed $\phi_1-\phi_2$. This function diverges at $t=t_*$ shown in \eqref{eq:cosmo_horizon}, which is the cosmological horizon for a static observer sitting at $\phi = (\phi_1+\phi_2)/2$, and turns negative for $t>t_*$. Intuitively, this is because the subsystem $A$ is timelike separated with itself in this case. }
    \label{fig:2pt_Func_real_time}
\end{figure}

\begin{figure}[h]  
    \centering
    \includegraphics[width=1.0\linewidth]{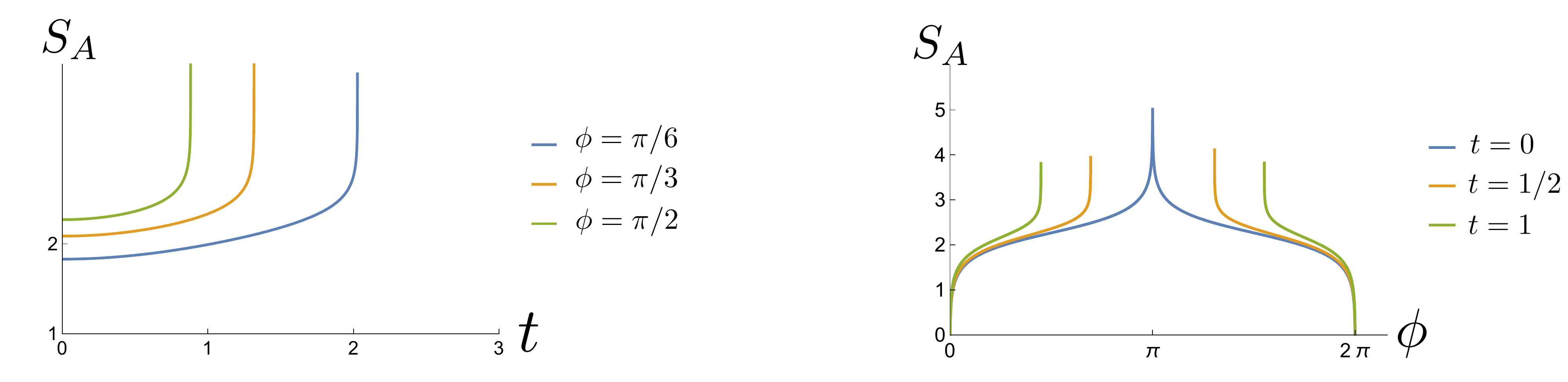}
    \caption{The left panel shows the entanglement entropy as a function of $t$ for fixed $|\phi_2-\phi_1|$. 
    The right panel shows the entanglement entropy as a function of $|\phi_2-\phi_1|$ for a fixed time $t$. 
    Both figures are restricted to $t<t_*$.
    The entanglement entropy diverges at $t=t_*$ shown in \eqref{eq:cosmo_horizon}, which is the cosmological horizon for a static observer sitting at $\phi = (\phi_1+\phi_2)/2$. Here we have set $\varepsilon_{\rm UV}=4\exp(-15/2)\approx 0.002$.}
    \label{fig:Entanglement entropy co-expanding segment}
\end{figure}

Let us then consider a subsystem $A$ with a fixed coordinate size $|\phi_2-\phi_1|\in[0,\pi)$ but not its proper length, on the same $t_1=t_2=t$ slice. This represents a segment co-expanding with the universe. Performing $\theta \rightarrow it$ to \eqref{eq:equal_theta}, we have
\begin{align}\label{eq:EE_equal_time}
    S_A
    =\frac{1}{6}\log\left(\frac{16\sin^2(\frac{\phi_1-\phi_2}{2}) \cosh^2(t)}{\varepsilon_{\rm UV}^2\left(3+2 \cos(\phi_1-\phi_2)\cosh^2(t)-\cosh(2t)\right)}\right).
\end{align}
Let us look at the argument of this formula and define
\begin{align}
    f(\phi_1-\phi_2,t) \equiv \frac{\sin^2(\frac{\phi_1-\phi_2}{2}) \cosh^2(t)}{\left(3+2 \cos(\phi_1-\phi_2)\cosh^2(t)-\cosh(2t)\right)}. 
\end{align}
A plot of this function is shown in figure \ref{fig:2pt_Func_real_time}. 
This function diverges when 
\begin{align}\label{eq:cosmo_horizon}
    t = t_*~~{\rm s.t.}~~e^{t} = \text{cot}\left(\frac{|\phi_2-\phi_1|}{4}\right)
\end{align}
This $t = t_*$ is exactly the timing when $A$ cross the ``cosmological horizon" of the static observer sitting at $\phi = (\phi_1+\phi_2)/2$. Refer figure \ref{fig:dS2}. Initially, $f(\phi_1-\phi_2,t)$ monotonically increases till $t=t_*$. After $A$ cross the ``cosmological horizon" $t > t_*$, $f(\phi_1-\phi_2,t)$ becomes negative.
This is because, after $t > t_*$, the subsystem $A$ will not be able to be accommodated in any static patch, 
and $A$ becomes timelike separated from itself. This is very similar to the analysis of timelike entanglement entropy \cite{DHMTT22,Narayan22,DHMTT23,MAP25}, where the entanglement entropy for a timelike subsystem becomes complex. Before $A$ cross the cosmological horizon $t<t_*$, \eqref{eq:EE_equal_time} still computes the entanglement entropy. See figure \ref{fig:Entanglement entropy co-expanding segment} for the time evolution of it in this time range.

\section{One-state Hilbert space and nontrivial observer-dependent ones}\label{sec:1D_space}

So far we have been focusing on the no-boundary density matrix realized by the Euclidean path integral on Euclidean elliptic dS, namely the real projective space. We have seen how the density matrix associated to local observers can be defined without defining the global no-boundary wave function. In this section, we comment on the direct canonical quantization in Lorentzian dS$/\mathbb{Z}_2$. We will take the free scalar theory, which has been previously discussed in \cite{Gibbons86,Sanchez86,PSV02}, as an example. We would like to note that there is no new technical constructions in this section --- all what we are doing here is to highlight some qualitative features from the standard canonical quantization and put it into the modern context of the study on the closed universe. This section may be regarded as a discussion section. 

We will emphasize the following aspect of the QFT in dS$/\mathbb{Z}_2$ --- the Hilbert space associated with the global spatial slice is one-dimensional, but an observer-dependent Hilbert space which has nontrivial dimensions can be constructed for each local observer. Physically, this is due to the fact dS$/\mathbb{Z}_2$ does not admit a global time orientation, and therefore cannot distinguish positive energy and negative, or creation operators and annihilation operators. Meanwhile, each static patch admits a well-defined time orientation. 

Let us first unfold the elliptic dS$_{d+1}$ to the standard global dS$_{d+1}$. A real scalar field profile $\varphi(t,\Omega_d)$ can be expanded by a set of mode functions $f_n(t,\Omega_d)$ as \cite{Allen85}
\begin{align}
    \varphi(t,\Omega_d) = \sum_n \left(a_n f_n(t,\Omega_d) + a_n^{*} f_n^*(t,\Omega_d)\right)~,
\end{align}
where each mode satisfies \cite{Allen85,PSV02}
\begin{align}\label{eq:mode_function}
    f_n(-t,\Omega_d^A) = f_n^*(t,\Omega_d^A)~. 
\end{align}
In the canonical quantization, we uplift $\varphi(t,\Omega_d)$ to a field operator 
\begin{align}
    \hat{\varphi}(t,\Omega_d) = \sum_n \left(\hat{a}_n f_n(t,\Omega_d) + \hat{a}_n^{\dagger} f_n^*(t,\Omega_d)\right)~,
\end{align}
where $\hat{a}_n$'s ($\hat{a}_n^{\dagger}$'s) are regarded as the annihilation (creation) operators. The Hilbert space of a scalar field in global dS is constructed by starting from a vacuum state $|\rm vac\rangle$ which is annihilated by the annihilation operators $\forall n, ~a_n|\rm vac\rangle = 0$, and then act the creation operators on it to generate new states. 
To be consistently defined on elliptic dS, $\hat{\varphi}(-t,\Omega_d^A) = \hat{\varphi}(t,\Omega_d)$ should be satisfied\footnote{We can also consider the antiperiodic boundary condition, which follows an essentially the same argument.}. Under \eqref{eq:mode_function}, we get
\begin{align}
    \hat{a}_n^\dagger = \hat{a}_n~. 
\end{align}
Therefore, if we start from a $|\rm vac\rangle$ and then try to construct the Hilbert space by acting the creation operators on it, $|\rm vac\rangle$ will be the only element in the Hilbert space since $|\rm vac\rangle$ is also annihilated by the creation operators. In this way, we end up with a one-dimensional Hilbert space. This peculiar feature originates from the fact that the elliptic dS itself does not admit a global time orientation. The creation operator is supposed to add ``positive" energy to the vacuum state. However, such a positivity cannot be globally defined because of the non-orientability. 

This observation also suggests that the standard construction of the Fock space should still work inside the static patch associated to each local observer, because it admits a consistent time orientation as discussed in section \ref{sec:causal_structure}. Indeed, given a choice of the static patch $\mathcal{C}$, one can define static-patch-dependent pairs of creation/annihilation operators $a^{\dagger\mathcal{C}}_n$, $a^{\mathcal{C}}_n$. Starting from a vacuum state which is annihilated by the annihilation operators, one can construct a nontrivial Fock space by acting creation operators on it \cite{PSV02}. 
The creation/annihilation operators used in this canonical quantization is observer-dependent, and so is the Hilbert space. Creation/annihilation operators relational to different static observers are connected by a Bogoliubov transformation, which is invertible. 

These features appearing in the canonical quantization are consistent with our study of the Euclidean QFT on $\mathbb{RP}^2$, where we cannot define a no-boundary wave function, but yet no-boundary density matrices associated with different observers. 

Recently, it has been argued that quantum gravity in a closed universe has a one-dimensional Hilbert space \cite{MM20,MV20,UWZ24} if one takes non-perturbative gravitational effects into consideration. Meanwhile, one can recover a nontrivial Hilbert space dimension by considering Hilbert spaces relational to observers \cite{AAIL25,HUZ25} (See also [\citenum{VChen25}, \citenum{NU25}, \citenum{Wei25}]). These features seem to be parallel to what we have discussed in dS$/\mathbb{Z}_2$, though they have distinct mathematical origins. Especially, the QFT in dS$/\mathbb{Z}_2$ does not involve any dynamical gravity --- it is all about QFT on a curved background. However, it is also difficult to say that gravity is not involved at all, since it is the causal structure in dS$/\mathbb{Z}_2$ which leads to this property, and the causal structure is also a key aspect of gravity. It would be interesting to understand to what extent the phenomenon we discussed in dS$/\mathbb{Z}_2$ can be regarded as ``nonperturbative phenomenon from perturbative physics"\footnote{We thank Yasunori Nomura for this comment.}. 

It is worth pointing out that the causal structure must be fine-tuned for this property to appear. If we instead considered a Lorentzian cylinder with flat metric quotient by a similar $\mathbb{Z}_2$ involution, then we will not be able to perform the canonical quantization even for each local observers, since there is no well-defined time directions associated to them. dS$/\mathbb{Z}_2$ is very much fine-tuned in this sense. It would be interesting to understand QFT in other types of spacetime with special causal structures. See \cite{MSSW24,MSW25} for some recent examples.

\section{Conclusion}\label{sec:conclusion}

Unlike global dS$_{d+1}$, where the Euclidean path integral on $\mathbb{S}^{d+1}$ prepares a no‑boundary wave function, the path integral on the Euclidean counterpart of dS$_{d+1}/\mathbb{Z}_2$, namely $\mathbb{RP}^{d+1}$, does not prepare a wave function. In section \ref{sec:EdS_NBDM} we proposed instead that the path integral over $\mathbb{RP}^{d+1}$ prepares a \emph{no‑boundary density matrix}. This is reminiscent of the ILM proposal \cite{ILM24} in the gravitational path integral, although our analysis is purely field‑theoretic. We characterized this no‑boundary density matrix by its entanglement entropy and R\'enyi entropies, and computed them explicitly in the two‑dimensional free Dirac fermion CFT. As explained in section~\ref{sec:replica_trick_CFT}, the availability of explicit bosonized expressions for both twist operators and crosscap states makes the analytic calculation possible.

We analytically studied the entanglement entropy of a subsystem $A$ in dS$_2/\mathbb{Z}_2$ in several setups (Section~\ref{sec:EERP2}). In particular, as $A$ approaches the entire spatial slice at $t=0$, the entropy diverges, indicating that the full slice is not associated with a pure state—consistent with the absence of a no‑boundary wave function. For a co‑expanding subsystem, the entropy exhibits a phase transition once $A$ grows beyond the scale at which it can be contained within any static patch and the density matrix interpretation collapses. 

We also revisited canonical quantization directly on dS$/\mathbb{Z}_2$. The standard global quantization scheme collapses on the full spatial slice, yielding a one‑dimensional (trivial) Hilbert space. By contrast, each static observer in dS$/\mathbb{Z}_2$ still possesses a nontrivial Hilbert space realized by the standard Fock construction.

Our work develops no‑boundary density matrices in elliptic de Sitter within a purely field‑theoretic framework, without dynamical gravity. We expect analogous configurations to contribute to the gravitational path integral that prepares the gravitational no‑boundary density matrix \cite{ILM24}; to our knowledge, this has not yet been realized. We leave a full gravitational implementation to future work.

\section*{Acknowledgements}
We are grateful to Stephano Antonini, Yuya Kusuki, Yasunori Nomura, Andy Strominger, Yasushi Yoneta and Mengyang Zhang for useful discussions. We would also like to thank Chen Bai for their kindful correspondences.
ZW is supported by the Society of Fellows at Harvard University.

\appendix
\section{2-point function of twist operators on non-orientable surfaces}\label{app:A}

In this appendix we summarize the result of the two-point function for twist operators in the free Dirac fermion CFT in the presence of crosscaps. Indeed, as explained in the bulk of the text, there are two different possible crosscaps states $\ket{{C}_+}$ and $\ket{{C}_-}$ of interest and twist operators can be interpreted as conformal boundary conditions \cite{OT14}, which can be either Neumann or Dirichlet. Here we summarize the different two-points functions that one can obtain in all the different cases, first for the Klein-bottle $\mathbb{K}^2$, and then for the real projective plane $\mathbb{RP}^2$.


\subsection{The \texorpdfstring{$\mathbb{K}^2$}{K2} case}
\label{sec:appendix two-point function KB}

We summarize all the different results for the two-point functions with the different choices mentioned in the bulk of the text. Here, we consider a Klein bottle parameterized by $\tau\in[0,2\alpha]$ and $l\in [0,2\pi)$ with $l\sim l+2\pi$. We use the complex coordinate $y \equiv \tau + il$.

\begin{itemize}
    \item \textbf{Neumann and ${C}_-$}
    \begin{align}
    &\bra{{C}}e^{-2\alpha H}\sigma^{(a)}(y_1,\bar{y}_1)\sigma^{(-a)}(y_2,\bar{y}_2)\ket{{C}} \label{eq:two-point function KB 1 good}= \sum_{w,\text{ even}}e^{-\frac{w^2 R^2}{\alpha'}\alpha}e^{\frac{R}{2 }\frac{a}{n}w\left(\left(y_1-y_2\right)+\left(\bar{y}_1-\bar{y}_2\right)\right)}\\
    &\times \left(\frac{\eta\left(\frac{2i\alpha}{\pi}\right)^6\theta_1\left(\frac{y_1+\bar{y}_2}{2\pi i}+\frac{i \pi}{2\pi i}\mid\frac{2i\alpha}{\pi}\right)\theta_1\left(\frac{y_2+\bar{y}_1}{2\pi i}+\frac{i \pi}{2\pi i}\mid\frac{2i\alpha}{\pi}\right)}{\theta_1\left(\frac{y_2+\bar{y}_2}{2\pi i}+\frac{i \pi}{2\pi i}\mid\frac{2i\alpha}{\pi}\right)\theta_1\left(\frac{y_1+\bar{y}_1}{2\pi i}+\frac{i \pi}{2\pi i}\mid\frac{2i\alpha}{\pi}\right)\theta_1\left(\frac{y_2-y_1}{2\pi i}\mid\frac{2i\alpha}{\pi}\right)\theta_1\left(\frac{\bar{y}_2-\bar{y}_1}{2\pi i}\mid\frac{2i\alpha}{\pi}\right)}\right)^{\left(\frac{ \,a\alpha'}{2 n}\right)^2}\,. \nonumber
\end{align}

    \item \textbf{Neumann and ${C}_+$}
\begin{align}
    &\bra{{C}}e^{-2\alpha H}\sigma^{(a)}(y_1,\bar{y}_1)\sigma^{(-a)}(y_2,\bar{y}_2)\ket{{C}}=\sum_{m,\text{ even}}e^{-\frac{ \alpha'm^2}{R^2}\alpha}e^{\frac{m}{R}\frac{a}{n}\left(\left(y_1-y_2\right)+\left(\bar{y}_1-\bar{y}_2\right)\right)}\\
    &\times \left(\frac{\eta\left(\frac{2i\alpha}{\pi}\right)^6\theta_1\left(\frac{y_2+\bar{y}_2}{2\pi i}+\frac{i \pi}{2\pi i}\mid\frac{2i\alpha}{\pi}\right)\theta_1\left(\frac{y_1+\bar{y}_1}{2\pi i}+\frac{i \pi}{2\pi i}\mid\frac{2i\alpha}{\pi}\right)}{\theta_1\left(\frac{y_2-y_1}{2\pi i}\mid\frac{2i\alpha}{\pi}\right)\theta_1\left(\frac{\bar{y}_2-\bar{y}_1}{2\pi i}\mid\frac{2i\alpha}{\pi}\right)\theta_1\left(\frac{y_1+\bar{y}_2}{2\pi i}+\frac{i \pi}{2\pi i}\mid\frac{2i\alpha}{\pi}\right)\theta_1\left(\frac{y_2+\bar{y}_1}{2\pi i}+\frac{i \pi}{2\pi i}\mid\frac{2i\alpha}{\pi}\right)}\right)^{\left(\frac{ \,a\alpha'}{2 n}\right)^2}\,. \nonumber
\end{align}
    \item  \textbf{Dirichlet and ${C}_-$}
    \begin{align}
    &\bra{{C}}e^{-2\alpha H}\sigma^{(a)}(y_1,\bar{y}_1)\sigma^{(-a)}(y_2,\bar{y}_2)\ket{{C}}=\sum_{w,\text{ even}}e^{-\frac{w^2 R^2}{\alpha'}\alpha}e^{\frac{R}{2 }\frac{a}{n}w\left(\left(y_1-y_2\right)+\left(\bar{y}_1-\bar{y}_2\right)\right)}\\
    &\times \left(\frac{\eta\left(\frac{2i\alpha}{\pi}\right)^6\theta_1\left(\frac{y_2+\bar{y}_2}{2\pi i}+\frac{i \pi}{2\pi i}\mid\frac{2i\alpha}{\pi}\right)\theta_1\left(\frac{y_1+\bar{y}_1}{2\pi i}+\frac{i \pi}{2\pi i}\mid\frac{2i\alpha}{\pi}\right)}{\theta_1\left(\frac{y_2-y_1}{2\pi i}\mid\frac{2i\alpha}{\pi}\right)\theta_1\left(\frac{\bar{y}_2-\bar{y}_1}{2\pi i}\mid\frac{2i\alpha}{\pi}\right)\theta_1\left(\frac{y_1+\bar{y}_2}{2\pi i}+\frac{i \pi}{2\pi i}\mid\frac{2i\alpha}{\pi}\right)\theta_1\left(\frac{y_2+\bar{y}_1}{2\pi i}+\frac{i \pi}{2\pi i}\mid\frac{2i\alpha}{\pi}\right)}\right)^{\left(\frac{ \,a\alpha'}{2 n}\right)^2}\,. \nonumber
\end{align}
    \item \textbf{Dirichlet and ${C}_+$}
\begin{align}
    &\bra{{C}}e^{-2\alpha H}\sigma^{(a)}(y_1,\bar{y}_1)\sigma^{(-a)}(y_2,\bar{y}_2)\ket{{C}} \label{eq:two-point function KB 2 good} =\sum_{m,\text{ even}}e^{-\frac{ \alpha'm^2}{R^2}\alpha}e^{\frac{m}{R}\frac{a}{n}\left(\left(y_1-y_2\right)+\left(\bar{y}_1-\bar{y}_2\right)\right)}\\
    &\times\left(\frac{\eta\left(\frac{2i\alpha}{\pi}\right)^6\theta_1\left(\frac{y_1+\bar{y}_2}{2\pi i}+\frac{i \pi}{2\pi i}\mid\frac{2i\alpha}{\pi}\right)\theta_1\left(\frac{y_2+\bar{y}_1}{2\pi i}+\frac{i \pi}{2\pi i}\mid\frac{2i\alpha}{\pi}\right)}{\theta_1\left(\frac{y_2+\bar{y}_2}{2\pi i}+\frac{i \pi}{2\pi i}\mid\frac{2i\alpha}{\pi}\right)\theta_1\left(\frac{y_1+\bar{y}_1}{2\pi i}+\frac{i \pi}{2\pi i}\mid\frac{2i\alpha}{\pi}\right)\theta_1\left(\frac{y_2-y_1}{2\pi i}\mid\frac{2i\alpha}{\pi}\right)\theta_1\left(\frac{\bar{y}_2-\bar{y}_1}{2\pi i}\mid\frac{2i\alpha}{\pi}\right)}\right)^{\left(\frac{ \,a\alpha'}{2 n}\right)^2}\,.\nonumber
\end{align}

\end{itemize}
In cases where we take $(y_1,y_2)$ at the same euclidean time, the first factor is equal to $\bra{{C}}e^{-2\alpha  H}\ket{C}$ and will be absorbed in the normalization. It is important to emphasize that, although all these two-point functions of twisted vertex operators are well-defined, not all of them are suitable for computing the entanglement entropy via the replica trick. Inserting a twist operator is equivalent to imposing a fermionic boundary condition \cite{OT14}, and the expression for the R\'enyi  entropy in \eqref{Rényi} as a product of independent two-point functions is valid only when the boundary condition of the crosscap state \eqref{eq:crosscap condition} (Neumann or Dirichlet) is compatible with that of the twist operator. Consequently, only \eqref{eq:two-point function KB 1 good} and \eqref{eq:two-point function KB 2 good} can be used to compute the R\'enyi  entropies. When we study the normalized two point functions:
\begin{equation}
    \frac{\bra{{C}}e^{-2\alpha H}\sigma^{(a)}(y_1,\bar{y}_1)\sigma^{(-a)}(y_2,\bar{y}_2)\ket{{C}}}{\bra{{C}}e^{-2\alpha H}\ket{{C}}}\,,
\end{equation}
the first factor can be reabsorbed and we have, using $\alpha'=2$ and $R=1$ in $\ket{{C}_+}$ case:
\begin{equation}
    \frac{\sum_{m,\text{ even}}e^{- 2m^2\alpha}e^{m\frac{a}{n}\left(\left(y_1-y_2\right)+\left(\bar{y}_1-\bar{y}_2\right)\right)}}{\sum_{m,\text{ even}}e^{- 2m^2\alpha}}=\frac{\theta_3(\frac{a}{n}\frac{1}{\pi i}\left(y_1-y_2+\bar{y}_1-\bar{y}_2\right)|\frac{8 i \alpha}{\pi})}{\theta_3(0|\frac{8 i \alpha}{\pi})}\,.
\end{equation}
while in the $\ket{{C}_-}$ case we obtain
\begin{equation}
    \frac{\sum_{m,\text{ even}}e^{- \frac{1}{2}m^2\alpha}e^{m\frac{a}{2n}\left(\left(y_1-y_2\right)+\left(\bar{y}_1-\bar{y}_2\right)\right)}}{\sum_{m,\text{ even}}e^{- \frac{1}{2}m^2\alpha}}=\frac{\theta_3(\frac{a}{n}\frac{1}{4\pi i}\left(y_1-y_2+\bar{y}_1-\bar{y}_2\right)|\frac{2 i \alpha}{\pi})}{\theta_3(0|\frac{2 i \alpha}{\pi})}\,.
\end{equation}

\subsection{The \texorpdfstring{$\mathbb{RP}^2$}{RP2} case}
\label{sec:appendix two-point function RP2}

As explained in section \ref{sec:EERP2}, if we take the $\alpha\rightarrow \infty$ in the Klein bottle setup, 

\begin{align}
    \lim_{\alpha\rightarrow \infty}\frac{\bra{C}e^{-2\alpha H}\sigma^{(a)}(\tau_1,l_1)\sigma^{(-a)}(\tau_2,l_2)\ket{C}}{\bra{C}e^{-2\alpha H}\ket{C}}\equiv \left<\sigma^{(a)}(\tau_1,l_1)\sigma^{(-a)}(\tau_2,l_2)\right>_{\rm half ~cylinder}\,.
    \label{eq:def 2 pt fct half cyl}
\end{align}
one of the crosscap is sent to infinity and, naively, we expect that we can effectively get the correlators sandwiched by the vacuum state and one crosscap state (up to a numerical pre-factor due to the normalization of \eqref{eq:def 2 pt fct half cyl}). As we will shortly see, this is not always the case. We recall that we took:
\begin{equation}
    (y,\bar{y}) = (\tau + il, \tau - il), 
\end{equation}
and 
\begin{equation}
    (z,\bar{z})= (e^y, e^{\bar{y}})= ( e^{\tau+il},e^{\tau-il})\,.
\end{equation}
Let us set $\alpha'=2$. Taking $\alpha \rightarrow \infty$ in the four cases listed out in the previous subsection and performing a conformal transformation to the flat $\mathbb{RP}^2$, we have 

\begin{itemize}
    \item \textbf{Neumann and $C_-$}
\begin{equation}
    \left<\sigma^{(a)}(z_1,\bar{z}_1)\sigma^{(-a)}(z_2,\bar{z}_2)\right>_{\rm flat~\mathbb{RP}^2}
    \stackrel{?}{=}
    \left(\frac{|z_2\bar{z}_1+1||z_1\bar{z}_2+1|}{|z_1-z_2|^2|z_1\bar{z}_1+1||z_2\bar{z}_2+1|}\right)^{\left(\frac{a}{n}\right)^2}\,.
    \label{eq:two point RP2 good 1}
\end{equation}
\item 
\textbf{Neumann and $C_+$}
\begin{equation}
    \left<\sigma^{(a)}(z_1,\bar{z}_1)\sigma^{(-a)}(z_2,\bar{z}_2)\right>_{\rm flat~\mathbb{RP}^2}\stackrel{?}{=}\left(\frac{|z_1\bar{z}_1+1||z_2\bar{z}_2+1|}{|z_1-z_2|^2|z_2\bar{z}_1+1||z_1\bar{z}_2+1|}\right)^{\left(\frac{a}{n}\right)^2}\,.
\end{equation}
\item
\textbf{Dirichlet and $C_-$}
\begin{equation}
    \left<\sigma^{(a)}(z_1,\bar{z}_1)\sigma^{(-a)}(z_2,\bar{z}_2)\right>_{\rm flat~\mathbb{RP}^2}\stackrel{?}{=}\left(\frac{|z_1\bar{z}_1+1||z_2\bar{z}_2+1|}{|z_1-z_2|^2|z_2\bar{z}_1+1||z_1\bar{z}_2+1|}\right)^{\left(\frac{a}{n}\right)^2}\,.
\end{equation}
\item
\textbf{Dirichlet and $C_+$}
\begin{equation}
    \left<\sigma^{(a)}(z_1,\bar{z}_1)\sigma^{(-a)}(z_2,\bar{z}_2)\right>_{\rm flat~\mathbb{RP}^2}\stackrel{?}{=}\left(\frac{|z_2\bar{z}_1+1||z_1\bar{z}_2+1|}{|z_1-z_2|^2|z_1\bar{z}_1+1||z_2\bar{z}_2+1|}\right)^{\left(\frac{a}{n}\right)^2}\,.
    \label{eq:two point RP2 good 2}
\end{equation}
\end{itemize}
Note that we did not assume $z_1,z_2$ have the same Euclidean time. Some of the results are incompatible to be the flat $\mathbb{RP}^2$ 2-point function since they do not respect the crossing symmetry on $\mathbb{RP}^2$. Indeed, up to a universal factor which depends on the conformal dimensions of the inserted operators, the two-point functions can only depend on a function of the moduli of $\mathbb{RP}^2$ with two punctures \cite{FPS93,Tsiares:2020ewp}:
\begin{equation}
    \eta= \frac{|z_1-z_2|^2}{\left(1+|z_1|^2\right)\left(1+|z_2|^2\right)}\,.
\end{equation}
Consequently the only sensible answer is:
\begin{equation}
    \left<\sigma^{(a)}(z_1,\bar{z}_1)\sigma^{(-a)}(z_2,\bar{z}_2)\right>_{\rm flat~\mathbb{RP}^2}=\left(\frac{|z_2\bar{z}_1+1||z_1\bar{z}_2+1|}{|z_1\bar{z}_1+1||z_2\bar{z}_2+1||z_1-z_2|^2}\right)^{\left(\frac{a}{n}\right)^2}\,.
\end{equation}
which is realized by either the $(\sigma^{(a,\,\pm)},\ket{{C}_\pm})$ pairs.

This is in stark contrast to the Klein bottle case, where all four two-point functions were well-defined, but only two could be used to compute R\'enyi entropies. The two-point functions on \texorpdfstring{$\mathbb{RP}^2$}{RP2} were obtained by sending one crosscap state to infinity, under the assumption that it projects the state onto the vacuum. This procedure, however, fails when the zero-mode boundary condition of the crosscap \eqref{eq:crosscap condition} is incompatible with that of the twist operator. In such cases, the vacuum does not appear in the intermediate channel between the twist operators and the crosscap, and the projection onto the vacuum is invalid, this leads to a failure of crossing symmetry. The correct two-point functions on \texorpdfstring{$\mathbb{RP}^2$}{RP2}  are given by \eqref{eq:two point RP2 good 1} and \eqref{eq:two point RP2 good 2}.

\section{Crosscap quench in free Dirac fermion CFT}
\label{app:quench}

In this appendix, we apply the results in the main text to study the entanglement entropy in the crosscap quench \cite{CCR24,WY24} in free Dirac fermion CFT.

In a CFT, the crosscap quench is a dynamical setup where the initial state is given by a regularized crosscap state $e^{-\beta H/{4}} |C\rangle$ and unitary time evolved with the CFT Hamiltonian \cite{WY24}: 
\begin{align}
    |\Psi(t)\rangle = \mathcal{N} e^{-itH} e^{-\beta H/4} |C\rangle. 
\end{align}
Here, $\mathcal{N} \equiv \langle C |e^{-\beta H/2}|C\rangle^{1/2}$ is the normalization factor. This setup admits a translational invariance and is hence a homogeneous global quench setup. However, different from standard global quenches which start from a non-thermal low entangled state (often a ground state of a gapped Hamiltonian) and evolve to a thermodynamic equilibrium \cite{CC05,WAKWWMY25}, in the crosscap quench, local subsystems are already in the thermal states at the initial stage. It is the non-local correlations which are highly structured at the initial stage, and get scrambled under the time evolution \cite{WY24}. 

The lattice analogue of the crosscap state is known as the entangled antipodal pair (EAP) states \cite{CY24,Yoneta24}. Given a periodic spin chain consisting $2N$ spins, the EAP state is given by:
\begin{align}\label{eq:EAP}
    |{\rm EAP}\rangle = \bigotimes_{i=1}^N |\Phi\rangle_{i,i+N}, 
\end{align}
where $|\Phi\rangle_{i,i+N}$ is a maximally entangled state between the $i$-th spin and the $(i+N)$-th spin such as $\left(|\uparrow\rangle_i\otimes|\uparrow\rangle_{i+N} + |\downarrow\rangle_i\otimes|\downarrow\rangle_{i+N}\right)/\sqrt{2}$. The lattice analogue of the crosscap quench has been analyzed in \cite{CCR24,WY24}. 

In the following, we work out the time evolution of the entanglement entropy for a subsystem under the crosscap quench analytically in the free Dirac fermion system. For simplicity, we set the spatial length of the CFT to be $2\pi$. 

In the following expressions we use $\alpha = \beta/4$ in eq.\eqref{eq:twist_2p} with $\beta$ having the interpretation of the inverse temperature of the system, the entanglement R\'enyi entropy of a single interval system $A = [0,l]$ turns out to be:
\begin{align}
    S^{(n)}_A &= \\
    &\frac{1}{12}\left(1+\frac{1}{n}\right) \log \left(\frac{1}{\varepsilon_{\rm UV}^2} \frac{\left|\theta_1 \left(\frac{l}{2\pi}|\frac{i\beta}{2\pi}\right)\right|^2 \cdot \left|\theta_1 \left(\frac{1}{\pi}\left(t+\frac{\pi}{2}-i\frac{\beta}{4}\right)|\frac{i\beta}{2\pi}\right)\right|^2}
    {\eta\left(\frac{i\beta}{2\pi}\right)^6 \cdot \left|\theta_1 \left(\frac{1}{\pi}\left(\frac{l}{2}+t+\frac{\pi}{2}-i\frac{\beta}{4}\right)|\frac{i\beta}{2\pi}\right)\right| 
    \cdot
    \left|\theta_1 \left(\frac{1}{\pi}\left(-\frac{l}{2}+t+\frac{\pi}{2}-i\frac{\beta}{4}\right)|\frac{i\beta}{2\pi}\right)\right|}\right)\,\nonumber 
\end{align}
where ${\varepsilon_{\rm UV}}$ is a UV cutoff corresponding to the latiice distance. 
Taking the $n\rightarrow1$ limit, we obtain the entanglement entropy for $A$ as 
\begin{align}
    S_A(t,l) &= \\
    &\frac{1}{6} \log \left(\frac{1}{\varepsilon_{\rm UV}^2} \frac{\left|\theta_1 \left(\frac{l}{2\pi}|\frac{i\beta}{2\pi}\right)\right|^2 \cdot \left|\theta_1 \left(\frac{1}{\pi}\left(t+\frac{\pi}{2}-i\frac{\beta}{4}\right)|\frac{i\beta}{2\pi}\right)\right|^2}
    {\eta\left(\frac{i\beta}{2\pi}\right)^6 \cdot \left|\theta_1 \left(\frac{1}{\pi}\left(\frac{l}{2}+t+\frac{\pi}{2}-i\frac{\beta}{4}\right)|\frac{i\beta}{2\pi}\right)\right| 
    \cdot
    \left|\theta_1 \left(\frac{1}{\pi}\left(-\frac{l}{2}+t+\frac{\pi}{2}-i\frac{\beta}{4}\right)|\frac{i\beta}{2\pi}\right)\right|}\right)\,.\nonumber 
\end{align}
This seemingly complicated formula has a simple behavior. In order to see this, let us look at some explicit examples and limits. 

\begin{figure}[h!]      \centering\includegraphics[width=16cm]{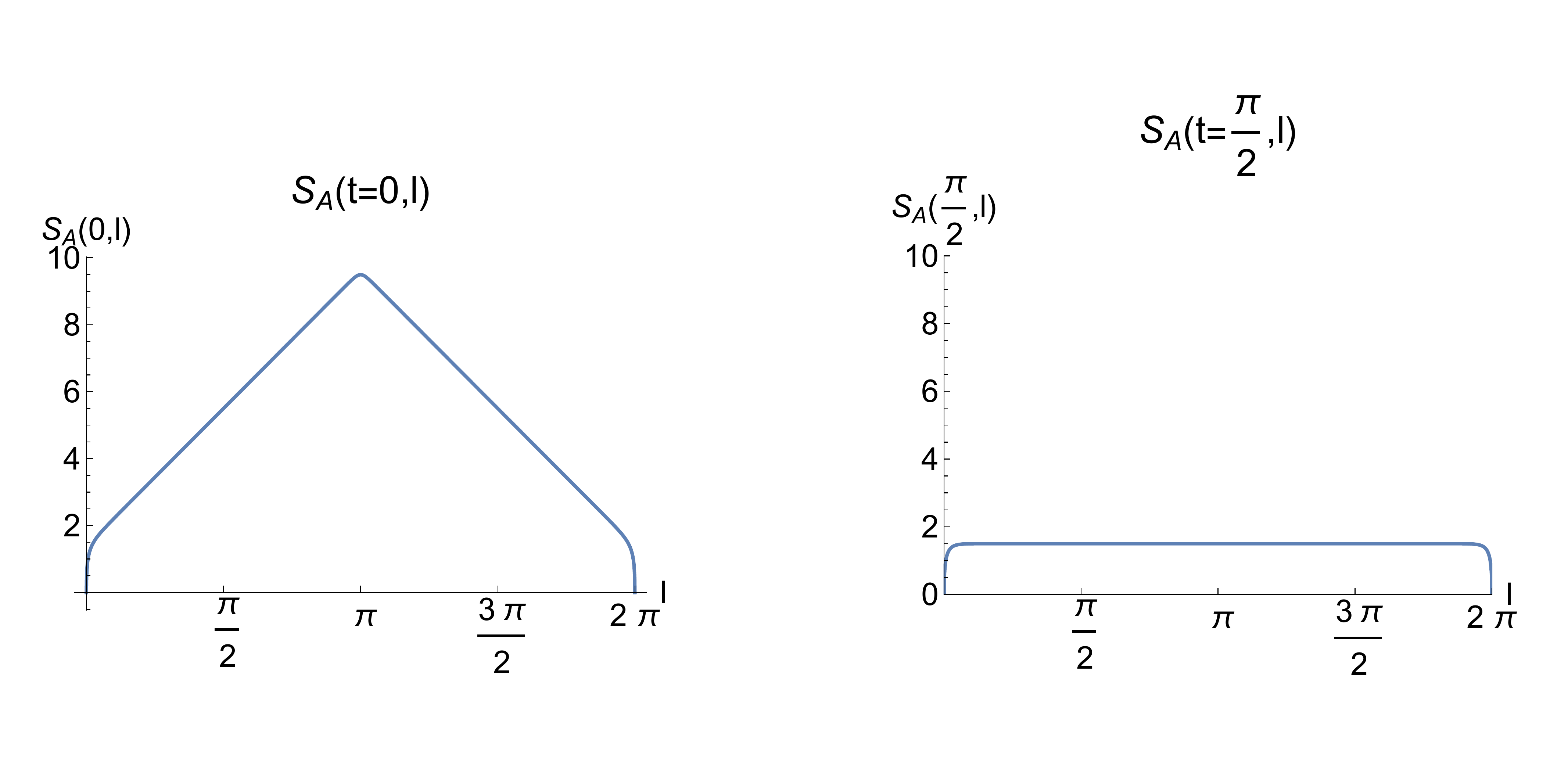}
    \caption{The entanglement entropy $S_A(t,l)$ of a single interval subsystem $A$ as a function of its length  $l$ at different but fixed time. The left panel shows $t=0$ while on the right panel $t=\frac{\pi}{2}$. We chose $\varepsilon_{\rm UV}=1$ and $\beta=\frac{1}{5}$. 
    For sufficiently large $l$, 
    the entanglement entropy does not scale with $l$ when $t=\frac{\pi}{2}$. By contrast, it grows linearly as a function of $l$ at $t=0$. }
    \label{fig:quench_initial}
\end{figure}

First of all, the left panel of figure~\ref{fig:quench_initial} shows the entanglement entropy of the initial state $S_{A}(0,l)$ as a function of $l$. We can see that $S_{A}(0,l)$ exhibits a Page curve-like behavior, where $S_{A}(0,l)$ increases  when $l$ becomes larger, reaches its maximum at $l = \pi$ and then decreases in a symmetric way. Especially, at the high temperature limit $\beta \rightarrow 0$, the entanglement entropy exhibits a volume law, 
\begin{align}
    S_A(0,l) = \frac{\pi l}{3\beta} + \cdots, 
\end{align}
reproducing the universal behavior found in \cite{WY24} for general CFT at central charge $c=1$. This indicates that the conformal crosscap state serves as a CFT analogue of the EAP state, as claimed in \cite{WY24}, as opposed to that the conformal boundary state serves as a CFT analogue of the product state \cite{MRTW14}. On the other hand, the right panel of figure~\ref{fig:quench_initial} shows the entanglement entropy $S_{A}(\pi/2,l)$ at a special time $t=\pi/2$. We can see that the entanglement entropy does not scale with $l$ when it is sufficiently large compared to the inverse temperature $\beta$, very similar to the entanglement structure of the conformal boundary state \cite{MRTW14}. 

\begin{figure}[h!]      \centering\includegraphics[width=16cm]{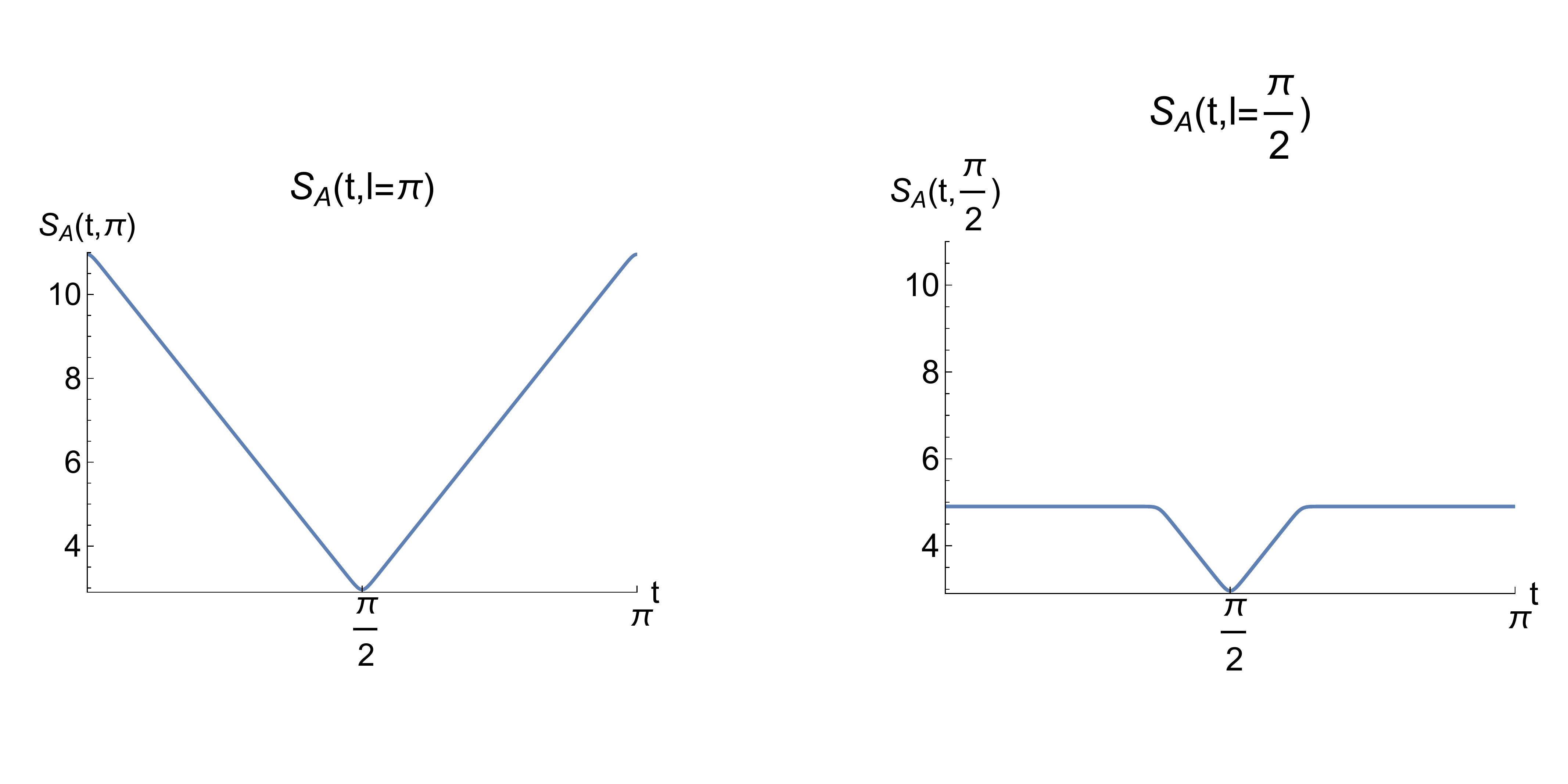}
    \caption{The entanglement entropy $S_A(t,l)$ of a single interval subsystem $A$ as a function of time $t$.  The left figure shows $l=\pi$, the case when $A$ is exactly a half of the whole system.
    The left panel shows the case of $l=\frac{\pi}{2}$. We chose $\varepsilon_{\rm UV}=1$ and $\beta=\frac{1}{5}$.}
    \label{fig:quench_half_system}
\end{figure}

Let us then look at the time evolution of the entanglement entropy. The right panel of figure \ref{fig:quench_half_system} shows the case of $l=\pi$, where $A$ is exactly a half of the whole system. We can see that once the time evolution is turned on, the entanglement entropy for the half system monotonically decrease. The entanglement entropy turns over and starts to increase at $t=\pi/2$. The increasing turns over again at $t=\pi$, and the entanglement entropy oscillates with the period $\pi$. For a generic single interval with length $0<l<\pi$, the time evolution is slightly more complicated. The left panel of figure\ref{fig:quench_half_system} shows such an example. We can see that the entanglement entropy stays at the initial thermal value till $t = (\pi-l)/2$, decreases, turns over at $t=\pi/2$, and returns to the thermal value at $t=(\pi+l)/2$. Especially, at the thermodynamic limit $l/\pi \ll 1$, the single interval will stay in the equilibrium, consistent with the universal behavior investigated in \cite{WY24}. 

\begin{figure}[h]      \centering\includegraphics[width=14cm]{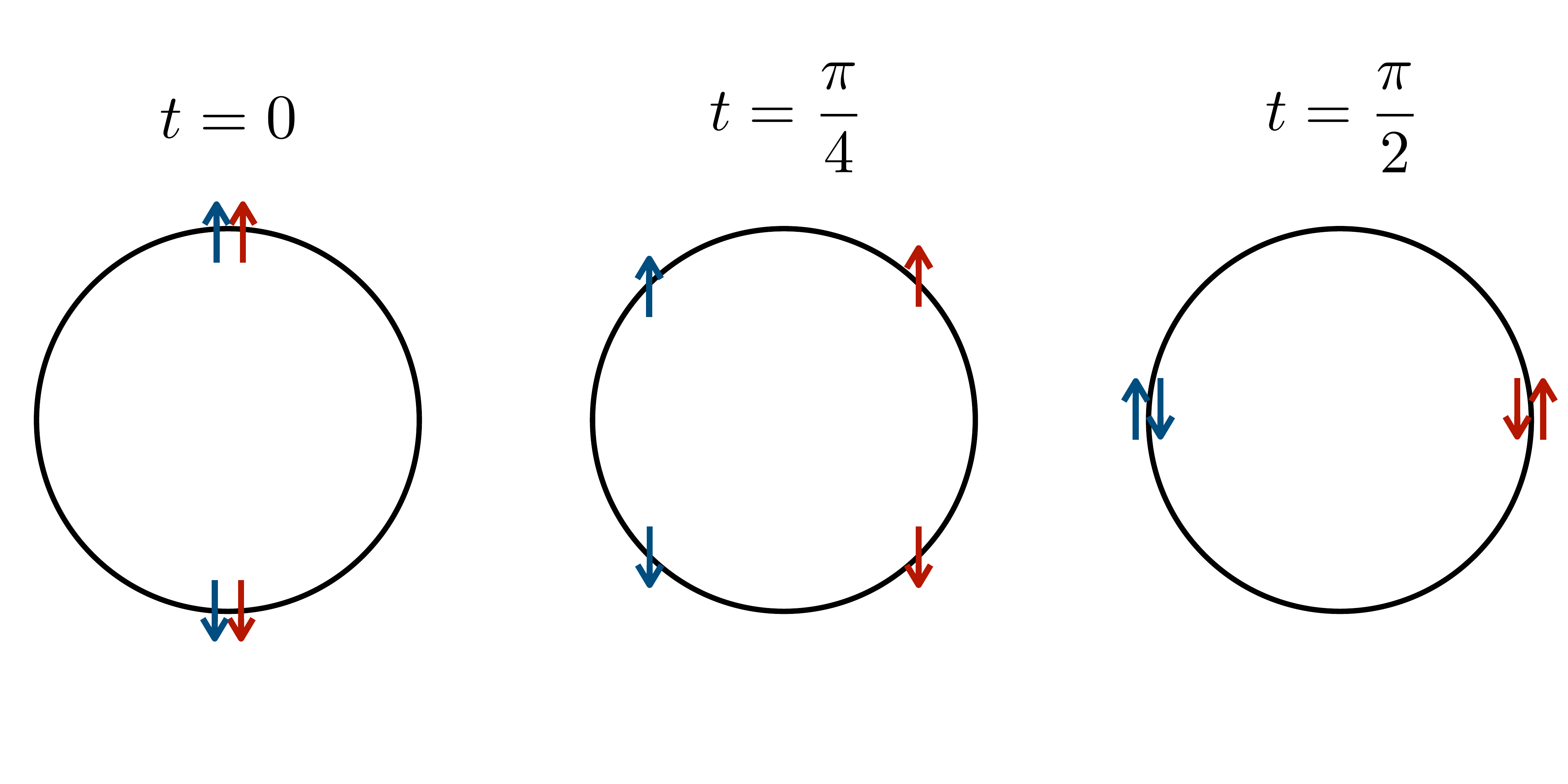}
    \caption{The quasi particle picture. The entangled particles are depicted as arrows with the same color, where the direction of the arrow does not carry meaning here. 
    Such particles exist at all the points along the circle, but only two (initial) antipodal pairs are shown. 
    At $t=0$ the two entangled quasi-particles are on opposite sides of the  circle. They move at the speed of light on opposite directions. The entanglement between two different spatial regions disappears at $t=\frac{\pi}{2}$ where the two quasi-particles of the same color are at the same place. This also corresponds to the minimum appearing in figure\ref{fig:quench_half_system} at $t=\frac{\pi}{2}$. Then the entanglement entropy increase again as the entangled quasi-particles get separated from each other. }
    \label{fig:quasi_particle}
\end{figure}

The time evolution behavior studied above can be schematically understood using a simple quasi-particle picture as follows. At the initial stage $t=0$, there are many entangled EPR pairs locating at the antipodal points $x$ and $x+\pi$, in a way similar to the EAP states \eqref{eq:EAP}. Once the time evolution is turned on, half of the EPR pairs move in the way such that the particles at $x$ move along the $+x$ direction and their entangled pairs at $x+l$ move along the $-x$ direction, both at the speed of light. At the same time, the other half of the EPR pairs move in the opposite way. figure~\ref{fig:quasi_particle} shows a sketch of this dynamics. 
The entanglement between $A$ and its complement are accounted for by the EPR pairs who sit across $A$ and its complement. For $0<l<\pi$, the number of such EPR pairs starts decreasing at $t= (\pi-l)/2$ and follows a time evolution qualitatively the same as the right panel of figure~\ref{fig:quench_half_system}. Especially, at $t=\pi/2$, any particle sits at the same location as its EPR pair, and the spatial entanglement disappears, qualitatively the same as the right panel of figure~\ref{fig:quench_initial}.

We end up with a comment on the relation to the global quench starting from a conformal boundary state \cite{CC05,TU10}. In the free Dirac fermion theory we have investigated here, the time evolution of the entanglement entropy after the crosscap quench has the same form as that of the boundary state quench \cite{TU10} with the time shifted by $\pi/2$. In fact, a conformal boundary state satisfies the constraint 
\begin{align}
    (L_n - \bar{L}_{-n})\ket{B} = 0
\end{align}
and it is straightforward to get
\begin{align}
    (L_n - (-1)^n\bar{L}_{-n})e^{-i\pi H/2}\ket{B} = 0. 
\end{align}
See e.g. \cite{Wei24} for a derivation. This means that $e^{-i\pi H/2}\ket{B}$ has the same symmetry as the crosscap state. This does not mean that $e^{-i\pi H/2}\ket{B}$ is a valid crosscap state satisfying the bootstrap equation, but we can see that this feature is manifested in the free Dirac fermion CFT.

\bibliographystyle{jhep}
\bibliography{DMindS}

\end{document}